\documentclass[11pt,double-line]{article}

\usepackage{setspace}

\usepackage[T1]{fontenc}
\usepackage{doi}
\usepackage{geometry}
\usepackage{cite}
\usepackage{amsmath}
\usepackage{url}
\usepackage{graphicx}
\usepackage{subcaption}
\usepackage{color}
\usepackage{soul}
\usepackage{dblfloatfix}    
\newcommand{\figref}[1]{\mbox{Figure~\ref{#1}}}

\usepackage{siunitx}
\usepackage[utf8]{inputenc}
 
\makeatletter\@ifundefined{date}{}{\date{}}
\makeatother

\evensidemargin\oddsidemargin \hoffset-22.4mm \voffset-28.4mm
\title{Achieving broadband directivity control with dual corona discharge transducers}
\author{Hervé Lissek, Rahim Vesal}
\date{Ecole Polytechnique Fédérale de Lausanne, Laboratory of Wave Engineering LWE\\ \today}

\begin{document}
\maketitle

\onehalfspacing

\begin{abstract}
Loudspeakers inherit their directivity from their geometry and dimensions. Enclosed loudspeakers are omnidirectional in the low frequency range, but their directivity depends on frequency for wavelengths smaller than the radiator size, precluding the directional control over the whole bandwidth. Loudspeakers pairs allow achieving simultaneously monopolar (in-phase) and dipolar (out-of-phase) sources, thus allowing directivity control. However, they are limited by their bulkiness, preventing extending controllable directivities over high frequencies. 

The Corona Discharge transducer (CDT) concept relies on ionizing an ultra-thin layer of air and oscillating it through an alternating electric field, generating sound without resorting to a mechanical membrane. This transducer combines a monopolar source linked to heat exchanges, and a dipolar linked to electrostatic forces, although these two sources strengths are interconnected, yielding a given unidirectional directivity.

In this paper, we propose to leverage the combination of monopole and dipole at the heart of the CDT concept to achieve controllable directivities by stacking two independent CDTs. The very thin dimensions of the CDT allows achieving coincident controllable monopolar and dipolar sound sources making the control of directivity over the whole operating frequency range. An analytical model of the dual CDTs concept is first compared to full-wave simulations, and an experimental prototype is finally assessed in anechoic conditions. Our findings opens the way to a new range of broadband directionally-controllable transducers that have application to sound generation, active noise reduction, or even non-reciprocal active acoustic metamaterials.
\end{abstract}

\section{Introduction}
The development of techniques aiming at controlling the directivity of electroacoustic transducers is conventionally driven by (spatial) audio motivations \cite{Rumsey2012}, such as improving sound coverage over a targeted zone \cite{Beranek1954, Kuttruff2024}, spatially localizing sound sources \cite{Chiariotti2019, Marmaroli2013} or avoiding  back reflections in sound reinforcement contexts \cite{Knaapen2014}. Similar techniques are also deployed in active noise cancellation contexts, either to separate the upstream and downstream sound fields on the sensors side \cite{Xiao2023}, or to directionally steer noise reduction performance on the actuators side \cite{Roure2006, Chen2011, Tanaka2017, Karkar2019, Debono2025}. Furthermore, controlling the directivity of sound sources is particularly timely in the context of active metasurfaces achieving anomalous refraction of sound \cite{Tan2024, Lissek2018}, active non-reciprocal metamaterials \cite{Popa2013, Tang2023, Guo2025}, and even at the heart of the Willis coupling which has attracted a significant attention in recent years \cite{Muhlestein2017, Groby2021, Demir2024, Chaplain2025}. Although  all directivity control techniques generally apply to both sensors (microphones) and actuators (loudspeakers), the focus will be put on loudspeaker systems in the following.

Loudspeakers directivities  are inherently dependent on frequency. Generally a loudspeaker system (driver + cabinet) is considered almost omnidirectional for wavelengths that are larger than the largest dimension of the loudspeaker system. But as frequency increases above the $kd_{\text{max}}=1$ "diffraction limit" (where $d_{\text{max}}$ is the largest dimension and $k$ the wavenumber), diffraction enters into play and yields frequency (relative to size)-dependent directivity patterns, that are inherently uncontrollable (as fixed by design) \cite{MorseIngard}. That means a given construction inevitably yields a given directivity.  Besides resorting to additional waveguides and horn radiators, likely to increase the loudspeaker systems dimensions \cite{Eargle1997}, several techniques have been developed in a view to controlling the directivity of electroacoustic transducers, provided the wavelength of operation is larger than the individual transducer dimensions. The most widespread delay-and-sum beamforming (DSB) consists in feeding a line (or surface) array of $N$ loudspeakers with time-delayed versions of a given audio signal in order to steer sound power towards a prescribed direction in space \cite{Eargle1997, Beer2016}. By deploying sub-arrays, it is even possible to extend the bandwidth of operation, but still limited in terms of frequency bandwidth of operation by the system dimensions \cite{Chen2014}. Parametric arrays leverage ultrasound focusing, relying on tiny ultrasound speakers (the reduced size of which allows covering a rather large frequency bandwidth), mapping the audible range by modulating two ultrasound signals of slightly different frequencies \cite{Shi2014}. Although performing in achieving highly directional sound beams, these techniques require a significant processing power. Therefore, much simpler directivity control configurations, such as gradient loudspeakers resorting on only a pair of sound sources, allow achieving unidirectional directivities \cite{Olson1973} although still governed by the diffraction limit. The following will rely on the latter technique as a simple use-case of directivity control with Corona Discharge Transducers (CDTs).

The CDT reported in Ref. \cite{Sergeev2020} consists of an ultra-thin layer of partially-ionized air particles that is put in motion within an intense, oscillating electric field, allowing moving the surrounding medium and generating sound without resorting to a physical mechanical radiator. This transduction principle is characterized by a combination of two sound sources: one monopolar source resulting from local heat release of the corona discharge, and one dipolar source linked to the back and forth electrostatic force. Although the two sources are interdependent (and proportional to the driving voltage, with fixed relative strengths), the very thin dimension of the CDT makes them almost coincident, making this type of loudspeaker an almost frequency-independent unidirectional sound source along the axis normal to the radiating surface (however, its lateral geometry and dimensions inevitably yield a prescribed directivity). Furthermore, it is possible to stack two adjacent CDTs in order to control independently the monopolar heat source and the dipolar force source. The latter configuration then allows achieving interesting directivity patterns, still outperforming conventional loudspeakers owing to their very thinness.

This papers reports a concept of dual CDT allowing achieving controllable unidirectional patterns over a broad frequency range. In a first section, the fundamental principle of the CDT is reminded, followed by the formulation of the sound field generated by a dual CDT. Then, numerical simulations are provided, which are finally experimentally verified on a laboratory prototype. Concluding remarks lead to discussion on the applicability of the dual CDT to directional audio, as well as active noise control \cite{Sergeev2022b} and as a unit-cell for active metamaterials and metasurfaces \cite{Sergeev2023}.

\section{The Corona Discharge Transducer}
\subsection{Description of the transduction principle}\label{sec:description}

\begin{figure}[ht]     
\centering
 \includegraphics[width=7.8cm]{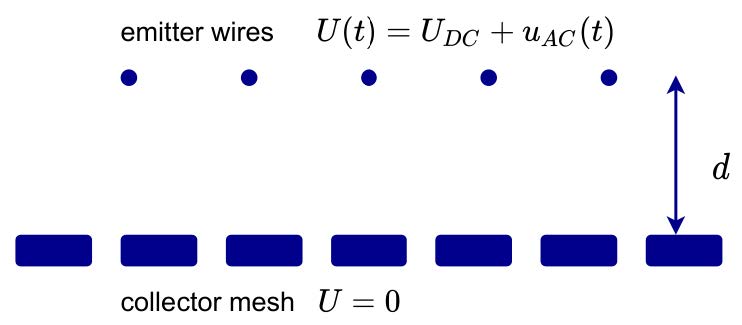}
 \caption{Schematic cut view of the electrode pair: the corona electrode consisting of an array of ultra-thin emitter wires parallel to each others (thin dots in this cut view), and a collector electrode consisting of a grounded perforated grid (dashed thick line in this cut view)}
\label{fig:CD geometry}
\end{figure}

Corona discharge transducers (CDTs) allow for sound generation thanks to the ionization of molecules in the surrounding fluid, moved by an intense oscillating electric field. In the "wire-to-grid" configuration, the transducer is constituted by a pair of electrodes (forming two parallel planes, as illustrated in the sketch of Figure \ref{fig:CD geometry}): the first one is a perforated metallic grid electrically connected to the ground, and is designated the "collector electrode"; the second one, designated "corona electrode", is made of an ultra thin wire (diameter of the order of 100 $\mu$m), arranged in a pattern of parallel lines along a same plane parallel and distant of $d$ to the collector plane, put at a sufficiently high voltage so as to trigger the extraction of electrons from surrounding molecules. The breakdown voltage, above which ionization occurs depending on the geometry of the CDT (inter-electrode distance, corona wires diameter) and on the medium properties (relative permittivity, density, temperature, etc.), is designated $U_0$ and is of the order of a few kVs in our study.

When applying an offset voltage $U_{DC}$ higher than $U_0$, the positive ions are accelerated from the ionization region (a few $\mu$m around the corona wires) towards the collector electrode. They then collide with the neutral particles present in the drift region (the inter-electrode space), which are then moved, giving rise to an "ionic wind" (constant flow). If an alternating voltage $u_{ac}(t)$ is superimposed to $U_{DC}$ (ensuring total voltage $U_{DC}+u_{ac}$ remains higher than the breakdown voltage), the transducer makes the surrounding fluid medium oscillate, responsible for sound generation.

The preceding studies \cite{Bequin2007, Sergeev2020, Sergeev2022a} have shown that this corona discharge transducer configuration could be modelled as a combination of two volumic sound sources, one monopolar "heat" source $\mathbf{H}$ (due to the heat release around the corona electrodes), and another dipolar "force" source $\mathbf{F}$ (relative to the electrostatic force moving the surrounding fluid back and forth), the orientation of which depends on the sign of the ac voltage: a positive $u_{ac}$ yields a force oriented towards the collector electrode and vice-versa.

An extremely simple electroacoustic model can be deduced from the coupling equations between the plasma generation, and the two sound sources strengths. Indeed, it has been shown that the offset voltage $U_{DC}$ can be linked to the current $I$ flowing through the corona wire  according to the Townsend formula:

\begin{equation}
I = C U_{DC}(U_{DC}-U_0)
\label{eq:Townsend}
\end{equation}

where $C$ is a constant that depends on the transducer geometry, and that can be experimentally determined. Note that Eq. \eqref{eq:Townsend} still holds for oscillating voltages, when substituting $U_{DC}+u_{ac}$ for $U_{DC}$. It is then possible to express the two sources strengths $\mathbf{H}$ and $\mathbf{F}$ as a function of the voltage feeding the transducer \cite{Sergeev2020, Sergeev2022a, Sergeev2023}. Assuming the ac voltage is much lower than the dc one, it is possible to linearize the expression of Eq. \eqref{eq:Townsend} and express  linear relationships between the heat source power  $\mathbf{H}$ (in \SI{}{W}) and bulk electrostatic force $\mathbf{F}$ (in \SI{}{N}) as:

\begin{equation}
\begin{aligned}
\mathbf{H}  &= C(3 U_{DC}^2-2U_{DC}U_0)u_{ac} \\
\vec{\mathbf{F}} & = \dfrac{C \vec{d}}{\mu_\text{ion}}(2U_{DC}-U_0)u_{ac} 
\end{aligned}
\label{eq:sources acoustiques}
\end{equation}
where $\mu_\text{ion}$ designates ions mobility, and $\vec{d}$ is the vector orthogonal to the electrode planes, defined from the corona electrode to the collector electrode ($d=||\vec{d}||$ the plasma thickness). 

It can be observed that the two sound sources are interdependent. Their individual influence is independent on frequency but on the constant parameters $U_0$, $U_{DC}$, $C$, $d$ and $\mu_\text{ion}$, and indirectly on the transducer cross-section area $S$. Also, the lateral dimensions of the CDT are likely to change the current-voltage law (Eq. \eqref{eq:Townsend}), thus $C$.

For the numerical study of Sec. \ref{sec:numerical-simulations}, we will define the heat source power density $h=\frac{H}{S.d}=h_0 u_{ac}$ (in $W/\text{m}^3$), and the dipolar force source density $f=\frac{||\vec{F}||}{S.d}=f_0 u_{ac}$ (in $N/\text{m}^3$), where $S$ is the transducer effective cross-section area.

\subsection{Sound field generated by a single CDT}\label{sec:single-CDT}

\begin{figure}[h!]
\begin{subfigure}[h]{0.55\linewidth}
       \centering
    \includegraphics[width=\linewidth]{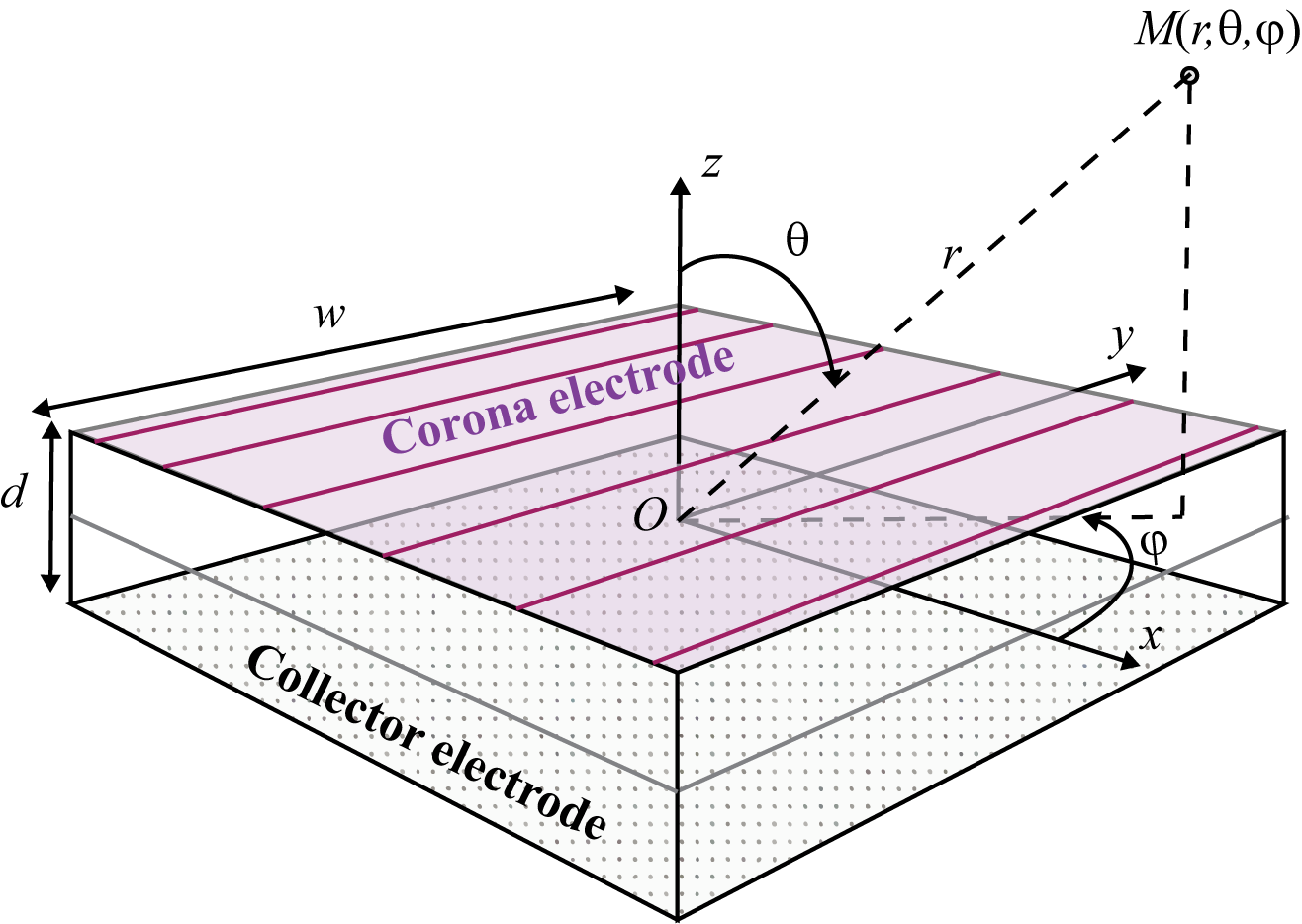}\caption{}
\end{subfigure}
\begin{subfigure}[h]{0.4\linewidth}
   \centering
    \includegraphics[width=0.8\linewidth]{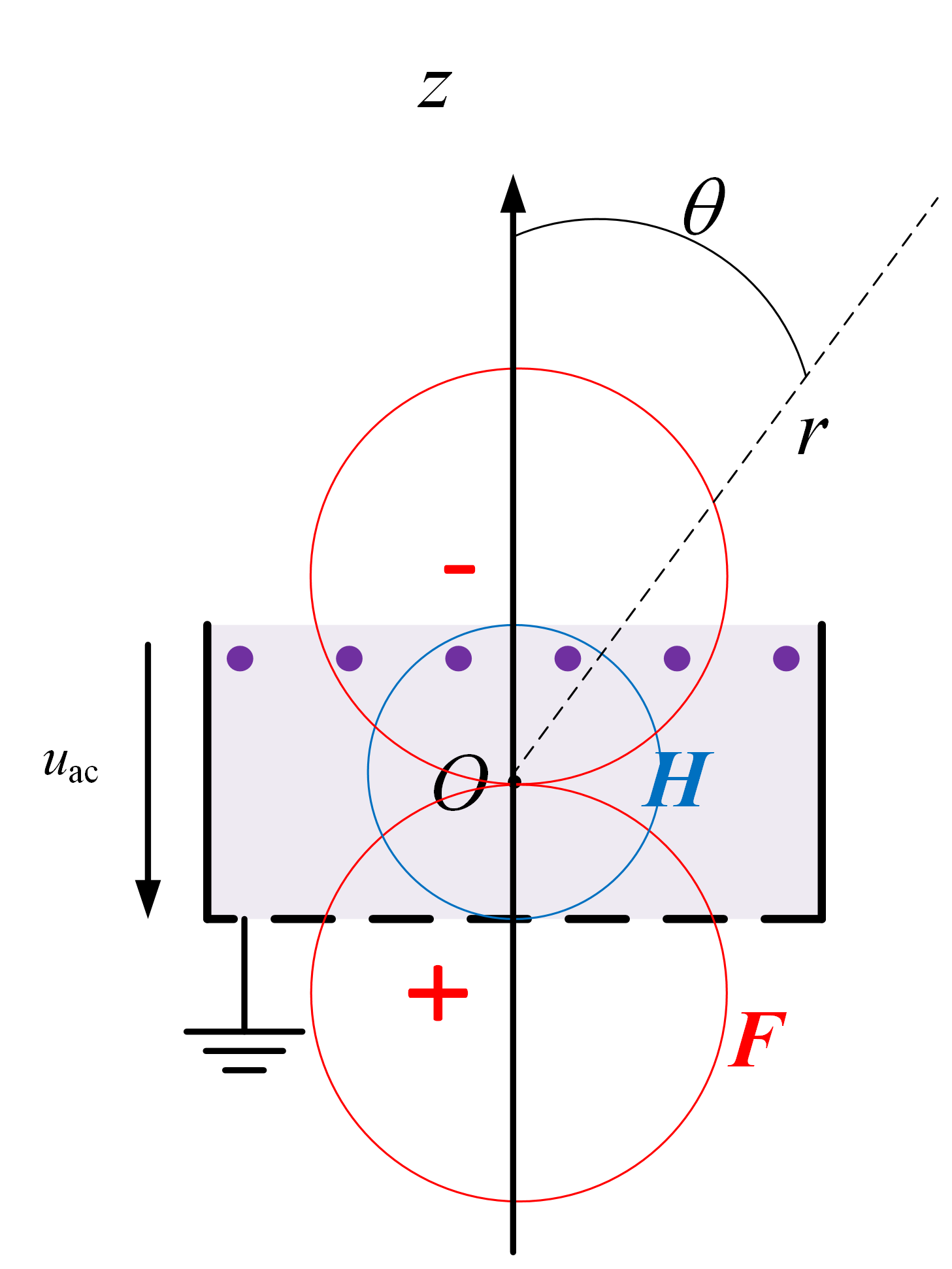}\caption{}
\end{subfigure}
    \caption{Sketch of a square CDT in free field: (a) 3D representation with cartesian and spherical coordinates definitions; (b) simplified cut view in the xOz plane. The blue circle represents the monopolar heat source, while the red circle pair highlights the dipolar force source, and their relative strength being qualitatively represented by the circles radii.}
    \label{fig:CDT-free field}
\end{figure}

Let's consider a square CDT of cross-section area $S=w^2$ where $w$ is the square width, and thickness $d$ (here the square geometry has been chosen for simplifying the fabrication of the prototype, but the following could also apply to any shape, for example a circular one). The corona electrode is made of a pattern of parallel thin wires (colored in deep purple), while the collector electrode is made of a perforated grid, illustrated with the black dots pattern. 
We will consider a spherical coordinates system ($\vec{r}, \theta,\varphi$), whose origin O is located at the CDT geometrical center, as illustrated on Figure \ref{fig:CDT-free field}. In this system, $r=\rVert \vec{r} \rVert$ is the distance to center O, angle $\theta$ designates the elevation with respect to $z$, and angle $\varphi$ designates the azimuth over the CDT reference plane (parallel to the electrodes planes comprising O). In addition to this coordinates system, we will denote $z$ the axis normal to the CDT reference plane, with unit vector $\vec{e}_z$. In this orientation, we can notice that $\vec{d} \cdot \vec{e}_z<0$, thus a negative sign in the following expressions of the electrostatic force. We will assume $d \ll \lambda $ where $\lambda$ denotes the wavelength, yielding the two sources centers (heat and force sources) are collocated at O. As $d$ is of the order of a \SI{5}{mm}, this assumption holds up to \SI{10}{kHz}. In this example, the height of corona electrode is supposed much smaller than the lateral dimension $w$ of the transducer: $d \ll w \ll \lambda$.

We can then derive the wave equation as \cite{Sergeev2020}:

\begin{equation}
\nabla^2p - \dfrac{1}{c^2}\dfrac{\partial ^2 p}{\partial t^2}=-\dfrac{\gamma -1 }{c^2}\dfrac{\partial h}{\partial t} + \vec{\nabla} . \vec{f}
\label{eq:wave equation time domain}
\end{equation}

where $\gamma$ is the specific heat ratio (equal to 1.4 in the air in adiabatic transformation). This can be formulated in the frequency domain as:
\begin{equation}
\left(\nabla^2 + k^2 \right)p=-\dfrac{j\omega(\gamma -1)}{c^2}h + \vec{\nabla}.\vec{f}
\label{eq:wave equation frequency domain}
\end{equation}
where $k$ is the wavenumber. 
By separating the "geometric" directivity $D_\text{piston}(\theta,\varphi)$ of the small piston of surface $S$ from the contributions of the heat and force sources distribution on directivity $D_\text{CDT}(\theta)$, the solutions of the wave equation in the frequency domain can be written as:
\begin{equation}
\begin{aligned}
p(\vec{r}) &= jk \left[\dfrac{(\gamma-1)}{c} \underbrace{\left(\iiint_Vh dV_0 \right)}_{\color{blue}H} + \operatorname{sgn}(\vec{d} \cdot \vec{e}_z)\left(\dfrac{j}{kr}+1 \right) \underbrace{\left( \iiint_V f dV_0\right)}_{\color{red}F=|| \vec{\mathbf{F}} ||}\cos \theta\right] D_{\text{piston}} \dfrac{e^{-jkr}}{4\pi r}\\
\end{aligned}
\label{eq:sound pressure solutions}
\end{equation}

where $\mathbf{H}$ and $\mathbf{F=\rVert \vec{F}  \rVert}$ are derived by linearization and expressed in Eq. \eqref{eq:sources acoustiques}. As the ratio between $\mathbf{H}$ and $\mathbf{F}$ is assumed constant for a given transducer and given DC voltage  $U_{DC}$, the individual contributions of the heat and force sources on the total sound pressure, thus directivity $D_\text{CDT}(\theta)$, are fixed by construction.

In the case of a square piston of width $w$, the "geometric" directivity function reads \cite{MorseIngard}:

\begin{equation}
D_{\text{piston}} (\theta,\varphi) =\mathbf{sinc}\left(\frac{kw}{2} \sin \theta\cos \varphi\right) \mathbf{sinc}\left(\frac{kw}{2} \sin \theta\sin \varphi\right) \\
\label{eq:piston directivity}
\end{equation}

where $\mathbf{sinc}$ designates the sine cardinal function. We will intentionally disregard the radiation contribution of the small piston of cross-section area $S$ to focus only on the heat and force source contributions $D_\text{CDT}(\theta)$ to the overall directivity. However, one should bare in mind that the directivities will be affected by the geometry of the radiating square piston at high frequencies, as soon as $\frac{kw}{2}>1$.


Let's assume now that the transducer dimensions are much smaller than the distance of observation ($r\gg w \gg d$), so that the problem can be simplified as the one of point sources in a free spherical domain. Also, discarding the directivity of the piston, we can simplify the problem to a 2D axisymmetric problem, so that the spherical coordinates can be limited to only $r$ and $\theta$ ($\varphi$ may only appear in $D_\text{piston}(\theta, \varphi)$).

In the far field ($kr \gg 1$), the sound pressure can be approximated as:
\begin{equation}
\begin{aligned}
p(r,\theta) &= jk\dfrac{e^{-jk r}}{4\pi r} \underbrace{ \left[ \dfrac{(\gamma-1)}{c} \mathbf{H} - \mathbf{F}\cos \theta \right]}_{\mathbf{F_\text{CDT}}}\\
\end{aligned}
\label{eq:far field p}
\end{equation}

The term inside the brackets of Eq. \eqref{eq:far field p} corresponds to the pressure forces exerted by the combination of the monopolar and dipolar sources at any angles $\theta$ over the boundaries of the dual CDT, that can be rewritten to take the form of a unidirectional function:

\begin{equation}
\begin{aligned}
\mathbf{F_\text{CDT}}(\theta) &= \mathbf{F_t} \underbrace{\left[(1-\mu) +\mu \cos (\theta+\pi)\right]}_{D_\text{CDT}(\theta) - \text{unidirectional}} \\
\end{aligned}
\label{eq:far field p cardioid}
\end{equation}

where $\mathbf{F_t}= \frac{(\gamma-1)\mathbf{H} +c\mathbf{F}}{c}$ is a total force (in \SI{}{N}, independent on frequency) and $\mu=\frac{c\mathbf{F}}{(\gamma-1)\mathbf{H}+c\mathbf{F}} \in [0,1]$ (dimensionless, independent on frequency) is the ratio of the bidirectional source to the overall directivity.

Then, with one single CDT, it is possible to achieve a given directivity $D_\text{CDT}(\theta)$, prescribed by the ratio between the Heat "weight" ($1-\mu$) and the Force "weight" ($\mu$), for any frequency. The following section will show how to decouple the monopole and dipole sources, and then fully control the directivity when two CDTs are stacked back-to-back.

\section{The Dual Corona Discharge Transducer}\label{sec:dual CDT}
\begin{figure}[ht]
    \centering
    \includegraphics[width=0.3\linewidth]{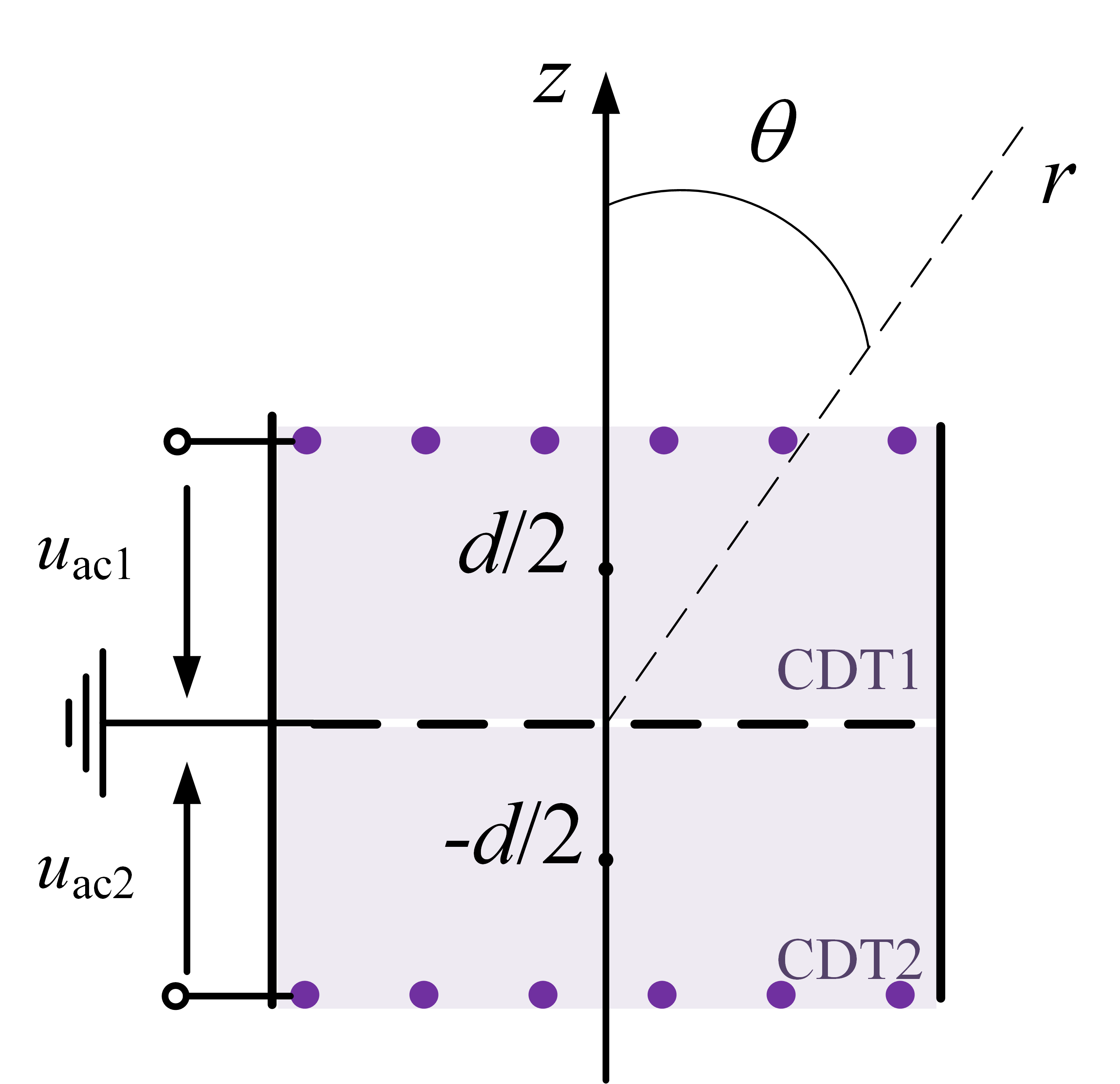}
    \caption{Simplified sketch of a dual Corona Discharge Transducer in free field (cut view along the xOz plane}
    \label{fig:dual CDT-free field}
\end{figure}

\subsection{General description}
Now we stack two CDTs with same shape and dimensions back-to-back, namely sharing a common collector electrode, as illustrated in Figure \ref{fig:dual CDT-free field}. For the sake of generality, we can consider that the two CDTs may present slight differences in terms of construction, thus giving rise to different electroacoustic parameters such as their coupling factors ($C_1$ and $C_2$), offset voltages ($U_{DC1}$ and $U_{DC2}$) or even breakdown voltages ($U_{01}$ and $U_{02}$). If they are individually driven by two different voltages $u_{ac1}$ and $u_{ac2}$, it will yield two different sets of monopolar sources $(\mathbf{H_1},\mathbf{H_2})$ and dipolar sources $(\mathbf{F_1},\mathbf{F_2})$. As in the preceding section, we will consider that the CDTs' thicknesses are still much smaller than the wavelength of interest (for example if $d=$\SI{5}{mm}, the total thickness is \SI{1}{cm}, thus subwavelength up to the range of 5 kHz). Then, following the same assumptions as in Sec. \ref{sec:single-CDT}, the total sound pressure in the far field becomes:

\begin{equation}
\begin{aligned}
p_t(r,\theta) &= jk  \left[ \left(\dfrac{\gamma -1}{c} \mathbf{H_1} - \mathbf{F_1} \cos \theta \right) \dfrac{e^{-jkr_1}}{4\pi r_1} + \left(\dfrac{\gamma -1}{c} \mathbf{H_2} + \mathbf{F_2} \cos \theta \right) \dfrac{e^{-jkr_2}}{4\pi r_2}\right]\\
\end{aligned}
\label{eq:far field dual CDT}
\end{equation}

where $r_1=|\vec{r}-\vec{r}_{01}|$ and $r_2=|\vec{r}-\vec{r}_{02}|$ are the distances of observations relative to the centers of each CDT $\vec{r}_{01}$ and $\vec{r}_{02}$. Without loss of generality, we can observe that the distance $d$ between the two centers is much smaller than the distance of observation $r_1 = r - \frac{d}{2} \cos \theta \approx r_2 = r +\frac{d}{2} \cos \theta \approx r \gg d$ and then:

\begin{equation}
\begin{aligned}
p_t(r,\theta) &= jk   \left(\dfrac{\gamma -1}{c} (\mathbf{H_1} +\mathbf{H_2} ) + (-\mathbf{F_1} + \mathbf{F_2})\cos \theta \right)  \dfrac{e^{-jkr}}{4\pi r}\\
& = jk \mathbf{F_t} \left(1-\mu_t +\mu_t \cos \theta \right)\dfrac{e^{-jkr}}{4\pi r}
\end{aligned}
\label{eq:far field dual CDT simplified}
\end{equation}

where $\mathbf{F_t}=\frac{\gamma-1}{c}(\mathbf{H_1}+\mathbf{H_2})+(\mathbf{F_2}-\mathbf{F_1})$ is a global force and $\mu_t$ is the unidirectional factor of the dual CDTs arrangement, that depend on $u_{ac1}$ and $u_{ac2}$ (and the parameters of the heat and force sources $U_{Dci}$, $U_{0i}$, $C_i$).

In the following, the directivity resulting from the monopole/dipole pair of the dual CDT (discarding $D_\text{piston}$) will be denoted:

\begin{equation}
    D_t(\theta)=(1-\mu_t) + \mu_t \cos \theta
    \label{eq:generic dual CDT directivity}
\end{equation}

\subsection{Case of ideally identical CDTs}\label{sec:ideal CDTs}
In the ideal case where both CDTs share exactly the same physical properties, then feeding each with a same ac voltage $u_{ac1}=u_{ac2}$ will cancel the dipolar terms (as the symmetry of the transducer yields opposite electrostatic force orientation on both sides of the collector electrode), thus achieving a fully monopolar source of amplitude $\frac{\gamma-1}{c}2\mathbf{H}$ with $\mathbf{H}=\mathbf{H_1}=\mathbf{H_2}$:

\begin{equation}
\begin{aligned}
p_t(r,\theta)|_{u_{ac1}=u_ {ac2}} &= jk   \dfrac{\gamma -1}{c} 2\mathbf{H}   \dfrac{e^{-jkr}}{4\pi r}\\
\end{aligned}
\label{eq:far field dual CDT monopolar}
\end{equation}

On the other hand, if the two CDTs are fed with opposite ac voltages ($u_{ac1}=-u_{ac2}$), this will cancel the monopolar contributions and thus yields a fully bi-directional source of amplitude $2\mathbf{F}$ with $\mathbf{F}=\mathbf{F_2}=-\mathbf{F_1}$:
\begin{equation}
\begin{aligned}
p_t(r,\theta)|_{u_{ac1}=-u_ {ac2}} &= jk  2 \mathbf{F}\cos \theta  \dfrac{e^{-jkr}}{4\pi r}\\
\end{aligned}
\label{eq:far field dual CDT dipolar}
\end{equation}

Finally, it is always possible to identify a combination of voltages $u_{ac1}$ and $u_{ac2}$ that allows achieving a targeted force $\mathbf{F_t}$ and directivity $D_t(\theta)=(1-\mu_t) + \mu_t\cos \theta$ after:

\begin{equation}
\left\{
\begin{aligned}
\mathbf{F_t}(1-\mu_t) &=  C(3 U_{DC}^2-U_{DC}U_0) \left(\dfrac{\gamma -1}{c}\right)(u_{ac1}+u_{ac2})\\
\mathbf{F_t}\mu_t & = \dfrac{C d}{\mu_\text{ion}}(2U_{DC}-U_0) (u_{ac2}-u_{ac1})
\end{aligned}
\right.
\label{eq:dual CDT equations}
\end{equation}

By denoting $\alpha_h=C\frac{\gamma -1}{c}(3U_{DC}^2-U_{DC}U_{0})$ and $\alpha_f=\frac{Cd}{\mu_\text{ion}}(2U_{DC}-U_0)$, the two sources coefficients, it finally yields:

\begin{equation}
\left\{
\begin{aligned}
u_{ac1} &=\mathbf{F_t} \dfrac{\alpha_f-(\alpha_h+\alpha_f)\mu_t}{2\alpha_h \alpha_f}  \\
u_{ac2} & =\mathbf{F_t}\dfrac{\alpha_f+(\alpha_h-\alpha_f)\mu_t}{2\alpha_h\alpha_f}
\end{aligned}
\right.
\label{eq: dual CDT voltages}
\end{equation}
thus $u_{ac1}$ and $u_{ac2}$ can be directly derived from the settings of $\mathbf{F_t}$ and $\mu_t$ and vice-versa. 
A more practical protocol consists in setting one voltage (eg. $u_{ac1}$) and prescribing a target unidirectional factor $\mu_t$, then it yields:
\begin{equation}
    \left\{
\begin{aligned}
\mathbf{F_t}= & 2\alpha_h \alpha_f\frac{1+\mu_t}{\alpha_f - (\alpha_h+\alpha_f)\mu_t}u_{ac1}\\
u_{ac2}= & \frac{\alpha_f+(\alpha_h-\alpha_f)\mu_t}{\alpha_f-(\alpha_h+\alpha_f)\mu_t} u_{ac1}
\end{aligned}
\right.
\label{eq: dual CDT Ft and uac2}
\end{equation}

\subsection{Case of CDTs with different characteristics}\label{sec:unknown CDTs}
In the most general case where the electroacoustic properties of the two CDTs are unknown, or if they happen to present slight geometrical and electrical differences, we can generalize the preceding protocol to derive the voltages $u_{ac1}$ and $u_{ac2}$ allowing achieving a targeted directivity $D_t (\theta)$ (and specifying for example one of the two ac voltages $u_{aci}$).

Let's consider the two CDTs (denoted with subscript $i$) characterized by the following parameters:

\begin{itemize}
    \item offset voltage $U_{DCi}$
    \item breakdown voltage $U_{0i}$
    \item cross section area $S_i$ (affecting the current-voltage constant $C_i$)
    \item inter-electrode gap $d_i$ (affecting the current-voltage constant $C_i$)
\end{itemize}

In this case, the heat and force sources coefficients read:

\begin{equation}
    \left\{
    \begin{aligned}
        \alpha_{h1} & = \dfrac{\gamma -1}{c}C_1(3 U_{DC1}^2-U_{DC1} U_{01}), & \alpha_{f1} & = \dfrac{C_1 d_1}{\mu_\text{ion}}(2 U_{DC1}-U_{01}) \\
        \alpha_{h2} & = \dfrac{\gamma -1}{c}C_2(3 U_{DC2}^2-U_{DC2} U_{02}) , & \alpha_{f2} & = \dfrac{C_2 d_2}{\mu_\text{ion}}(2 U_{DC2}-U_{02})     
    \end{aligned}
    \right.
    \label{eq: dual CDT unknown coefficients}
\end{equation}

There exists pairs of voltages $(u_{ac1,m},u_{ac2,m})$ for which the dual CDTs arrangement yields an omnidirectional sound source ($\mu_t=0$), and another pair $(u_{ac1,d},u_{ac2,d})$ for which the dual CDTs yields a bi-directional source ($\mu_t=1$). Experimentally, it is possible to identify such two pairs of voltages (see Sec. \ref{sec:exp calibration}) and deduce the relationships between $\alpha_{f1}$ and $\alpha_{f2}$ on the one hand, and between $\alpha_{h1}$ and $\alpha_{h2}$ on the other hand:

\begin{equation}
\left\{
    \begin{aligned}
        \alpha_{f2} &= \alpha_{f1} \dfrac{u_{ac1,m}}{u_{ac2,m}}\\
        \alpha_{h2} &= -\alpha_{h1} \dfrac{u_{ac1,d}}{u_{ac2,d}}
    \end{aligned}
    \right.
        \label{eq:dual CDT voltage-unknown}
\end{equation}

Also, computing the ratio between the sound pressures measured at a same distance $r$ and at angle $\theta$=\SI{0}{rad}, for both cases ($p_m(r,0)=p_t(r,0)|_{\mu_t=0}$ for the monopole setting and $p_d(r,0)=p_t(r,0)|_{\mu_t=1}$ for the dipole setting), yields the ratio between $\alpha_{h1}$ and $\alpha_{f1}$ as:

\begin{equation}
   \dfrac{\alpha_{h1}}{\alpha_{f1}}= \dfrac{u_{ac2,d}}{u_{ac2,m}} \dfrac{p_m(r,0)}{p_d(r,0)}
   \label{eq:alphah1/alphaf1}
\end{equation}

This ratio can finally be used to extract the voltage pair $(u_{ac1,\mu_t},u_{ac2,\mu_t})$ achieving the target directivity $\mu_t$ after Eq. \eqref{eq:dual CDT equations}:

\begin{equation}
\dfrac{1-\mu_t}{\mu_t}=\dfrac{\alpha_{h1}}{\alpha_{f1}} \dfrac{u_{ac2,m}}{u_{ac2,d}} \dfrac{u_{ac1,\mu_t} u_{ac2,d}-u_{ac1,d}u_{ac2,\mu_t}}{u_{ac1,m}.u_{ac2,\mu_t}-u_{ac1,\mu_t}.u_{ac2,m}}
    \label{eq:uac1mu and uac2mu}
\end{equation}

which finally yields:
\begin{equation}
\dfrac{u_{ac2,\mu_t}}{u_{ac1,\mu_t}}=\dfrac{\mu_t \frac{\alpha_{h1}}{\alpha_{f1}}+(1-\mu_t)}{\mu_t \frac{\alpha_{h1}}{\alpha_{f1}}\frac{u_{ac1,d}}{u_{ac2,d}}+(1-\mu_t) \frac{u_{ac1,m}}{u_{ac2,m}}  }
    \label{eq:uac2mu/uac1mu}
\end{equation}

In practice, we can fix one of the two CDT voltages (for example $u_{ac1}$) and keep the same value for all settings, in this case, $u_{ac1,m}=u_{ac1,d}=u_{ac1,\mu_t}=u_{ac1,\text{ref}}$, to simply derive $u_{ac2,\mu_t}$.

In the following, we will illustrate some examples of directivities achievable with the dual CDT assembly.

\section{Numerical simulations}\label{sec:numerical-simulations}
In this section, we will give some examples of directivities simulated  with COMSOL Multiphysics on a dual CDT arrangement, as described in Sec. \ref{sec:dual CDT}. In these simulation examples, we will consider two ideally identical CDTs, with constitutive parameters of Table \ref{tab:CDT parameters}, and follow the methodology described in Sec. \ref{sec:ideal CDTs}. Note that the electrical parameters ($C,U_0,U_{DC}$) provided in this table are estimated beforehand by measuring a prototype of same dimensions and geometry (here the  square transducer prototype presented in Sec. \ref{Sec:exp}).

\begin{table}
    \centering
    \begin{tabular}{lccc}
    \hline
    \textbf{Parameter}     &  \textbf{Symbol} & \textbf{Value}  & \textbf{Unit} \\
    \hline
      Square width   &  $w$ & 13 & cm \\
      CDT thickness   & $d$ &  5 & mm \\
      Current-voltage constant   & $C$ & 1.8 $10^{-11}$  & A/V$^2$\\
      Breakdown volatge   & $U_0$  & 6.16 & kV \\
      Offset voltage   & $U_{DC}$  & 8 & kV \\
      \hline
      Mass density of air   & $\rho$  & 1.2 & kg.m$^{-3}$ \\
      Speed of sound   & $c$ & 343 &  m.s$^{-1}$ \\
      Mobility of positive ions   & $\mu_\text{ion}$ &  1.1 $10^{-4}$ & m$^2$.V$^{-1}$.s$^{-1}$\\
      Specific heat ratio        & $\gamma$ & 1.4 & \\
      Temperature of air & $T_0$   &  293  & K \\
      Isobaric heat capacity &  $C_P$  & 1015  & J.kg$^{-1}$.K$^{-1}$ \\
         \hline
    \end{tabular}
    \caption{Parameters for the COMSOL model}
    \label{tab:CDT parameters}
\end{table}

\subsection{Model geometry}
\begin{figure}[ht]
\begin{subfigure}[h]{0.6\linewidth}
\centering
\includegraphics[width=\linewidth]{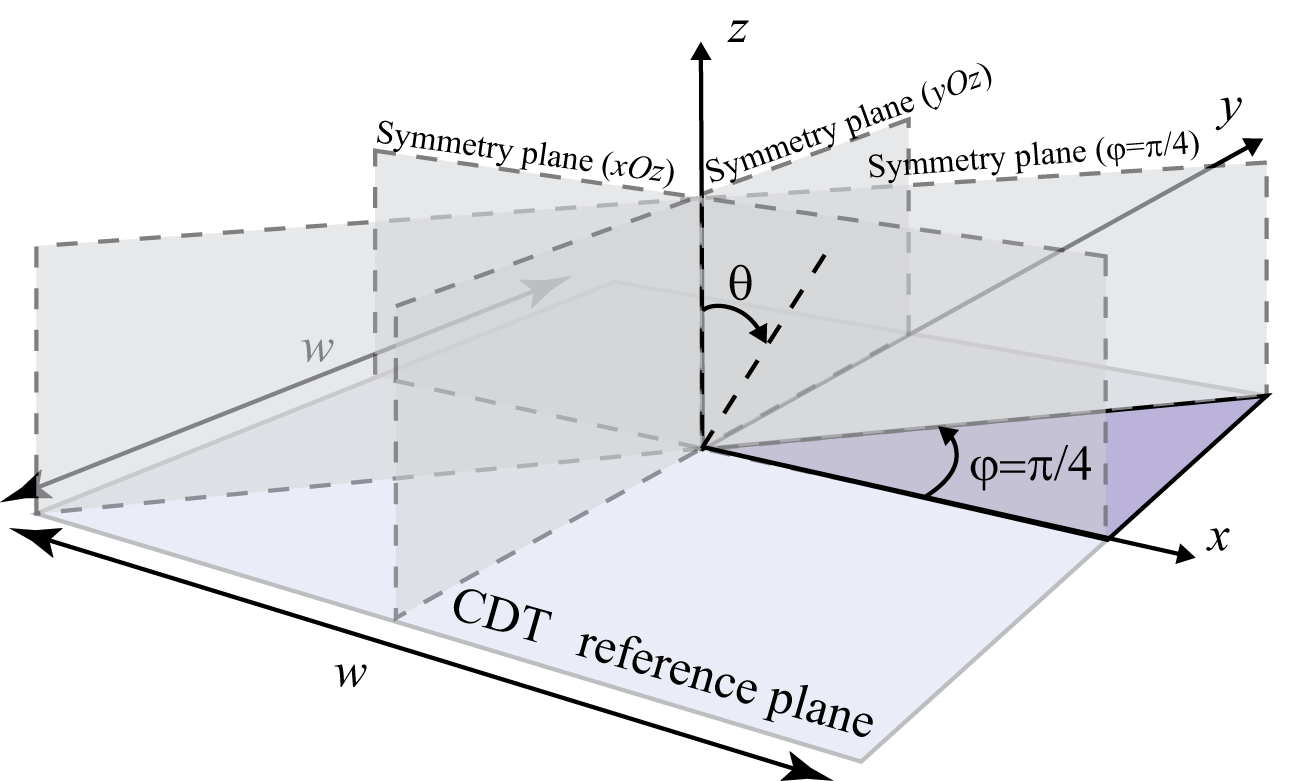}
\caption{}
\end{subfigure}
\hfill
\begin{subfigure}[h]{0.4\linewidth}
\centering
\includegraphics[width=\linewidth]{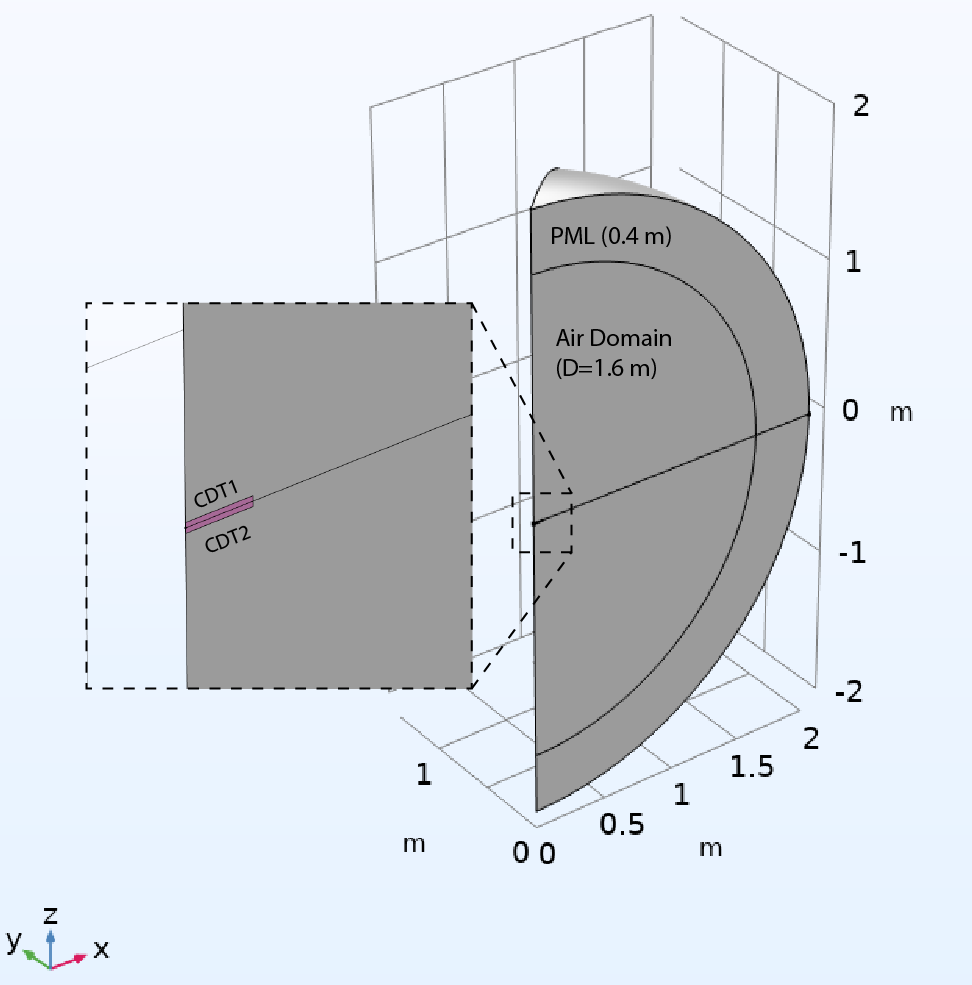}
\caption{}
\end{subfigure}%
\caption{Model geometry in COMSOL Multiphysics: (a) Sketch of the dual CDT reference plane $(z=0)$} highlighting symmetries, and definition of spherical coordinates angles $\theta$ and $\varphi$; (b) snapshot of the COMSOL 3D model simplified geometry.
\label{fig:COMSOL geometry}
\end{figure}

The numerical model is designed using COMSOL Multiphysics Acoustic Module in 3D space dimension. The fluid domain is defined as a sphere of radius $D=$ \SI{1.6}{m}, with a Perfectly Matched Layer (PML) of \SI{40}{cm}, making the simulation extent of \SI{2}{m}. Each of the individual CDT is modeled as identical parallelepipeds (Blocks) of square cross-section on the $xOy$ plane, with width $w$ and height $d$, sharing a common boundary at $z=0$. We assume here the $\varphi=0$ ($xOz$) and $\varphi=\frac{\pi}{2}$ ($yOz$), as well as the $\varphi=\frac{\pi}{4}$ planes, are symmetry planes, therefore simplifying the 3D domain comprised between planes $\varphi=0$ and $\varphi=\frac{\pi}{4}$. This space domain simplification is highlighted in \figref{fig:COMSOL geometry}(a) as a deep purple triangular surface over the reference plane $z=0$ (corresponding to the shared collector electrode). The corresponding simplified COMSOL 3D model is illustrated on \figref{fig:COMSOL geometry}(b).

\subsection{Model physics}\label{sec:model-physics}
The propagating domain, including the CDTs volume, is considered filled with air with mass density $\rho$ and speed of sound 
$c$.
Each CDT is modelled as the combination of a "Heat Source" of power density $Q_{heat,i}=h_0 u_{aci}$ where $h_0$ is defined in \ref{sec:description} (also specifying the ratio of specific heat capacities $\gamma$ and heat capacity at constant pressure $C_P$), and a "Dipolar Domain Source" of Dipole Source $q_d=(-1)^if_0 u_{aci}.\vec{e_z}$, where $i \in \{1,2\}$ designates the CDT number 1 or 2 (as in Figure \ref{fig:COMSOL geometry}). In this definition, the electrodes of the upper transducer CDT1 (resp. lower transducer CDT2) are arranged so that the projection of the electrostatic force $\vec{\mathbf{F}}_1$ (resp. $\vec{\mathbf{F}}_2$) over $z$ is negative (resp. positive) for a positive $u_{ac1}$ (resp. $u_{ac2}$), as in the geometry presented in Sec. \ref{sec:dual CDT}.

\subsection{Other settings and processing methodology}
The effect of the CDTs shape and dimensions on directivity start to enter into play for $\frac{kw}{2}>1$ where $w=$ \SI{13}{cm}, yielding a diffraction limit of approximately $f_{d}=$ \SI{1}{kHz}.
The study is performed in the frequency domain (acpr), for octave band frequencies spanning from $f_\text{min}=$\SI{125}{Hz} to $f_\text{max}=$\SI{1}{kHz}. The mesh inside the propagating domain and the CDTs is "Free Triangular" with maximal mesh size $\delta=\frac{c}{6. f_\text{max}}$, and the PML layer is meshed with a "Swept" mesh of maximum element size \SI{10}{cm}.

The processing starts by specifying voltage $u_{ac1}$ as well as the target directivity parameter $\mu_t$. From these values, $u_{ac2}$ is set after Eq. \eqref{eq: dual CDT Ft and uac2}, depending on the coefficients $\alpha_{h1}=\alpha_{h2}=\frac{\gamma -1}{c}C(3 U_{DC}^2-U_{DC} U_{0})$ and $\alpha_{f1}=\alpha_{f2}=\dfrac{C d}{\mu_\text{ion}}(2 U_{DC}-U_{0})$. The results are presented in the following subsection.

\subsection{Numerical results}
\subsubsection{Directivity}
The following figures show the achieved directivities, along the plane $\varphi=\frac{\pi}{4}$ for different parameter settings $\mu_t \in \{0,0.3,0.5,0.63,0.75,1\}$. The first series of curves (plain lines) represent the directivities computed for each octave band central frequency over the boundary of the propagating medium, namely the interior boundary of the PML, defined as $\text{abs}(acpr.p_t)/\text{maxop1}(\text{abs}(acpr.p_t))$, where the operator maxop1 processes the maximum values over this boundary. The second series (plus markers) represents the same quantity defined from the theoretical value $D_t(\theta)$ as in Eq. \eqref{eq:generic dual CDT directivity}, multiplied by the square piston directivity $D_\text{square}(\theta,\varphi)$ of Eq. \eqref{eq:piston directivity}.

We can observe that the dual CDT arrangement allows achieving the targeted directivites in each case (COMSOL and theoretical pressure fields perfectly match for each settings and each frequency). Also, we can observe that the square piston has a slight effect on directivity for $f>$ \SI{500}{Hz}, but can still considered omnidirectional up to \SI{1}{kHz} as $D_\text{square}(\theta= \frac{\pi}{2},\varphi=\frac{\pi}{4})_{|f=\SI{1000}{Hz}}=0.78 > \frac{1}{\sqrt{2}}$. 

\begin{figure}[h!]
\begin{subfigure}[h]{0.4\linewidth}
\centering
\includegraphics[width=0.9\linewidth]{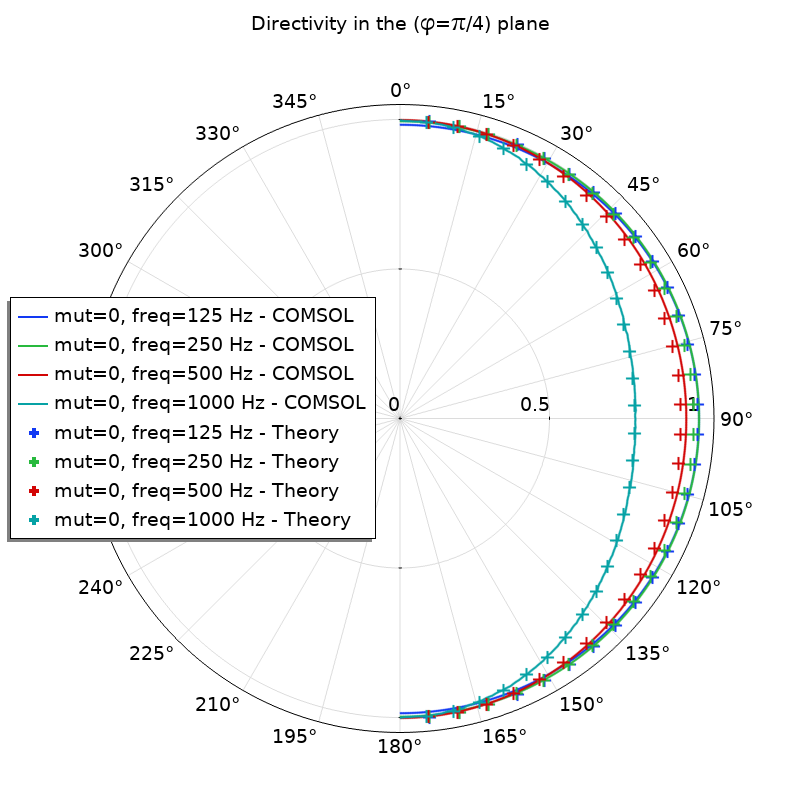}
\caption{}
\end{subfigure}
\hfill
\begin{subfigure}[h]{0.4\linewidth}
\centering
\includegraphics[width=0.9\linewidth]{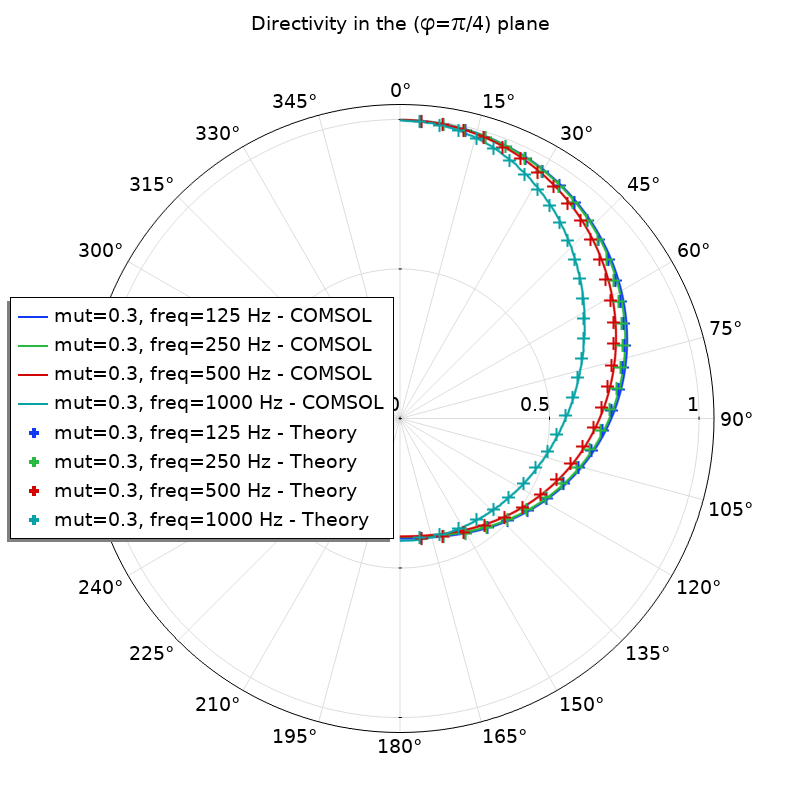}
\caption{}
\end{subfigure}

\begin{subfigure}[h]{0.4\linewidth}
\centering
\includegraphics[width=0.9\linewidth]{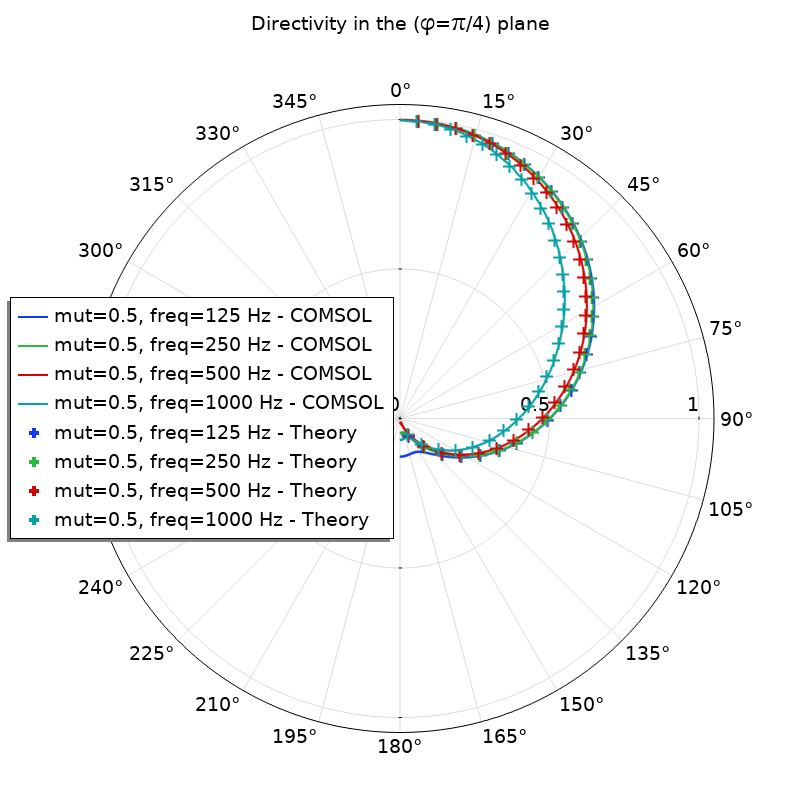}
\caption{}
\end{subfigure}%
\hfill
\begin{subfigure}[h]{0.4\linewidth}
\centering
\includegraphics[width=0.9\linewidth]{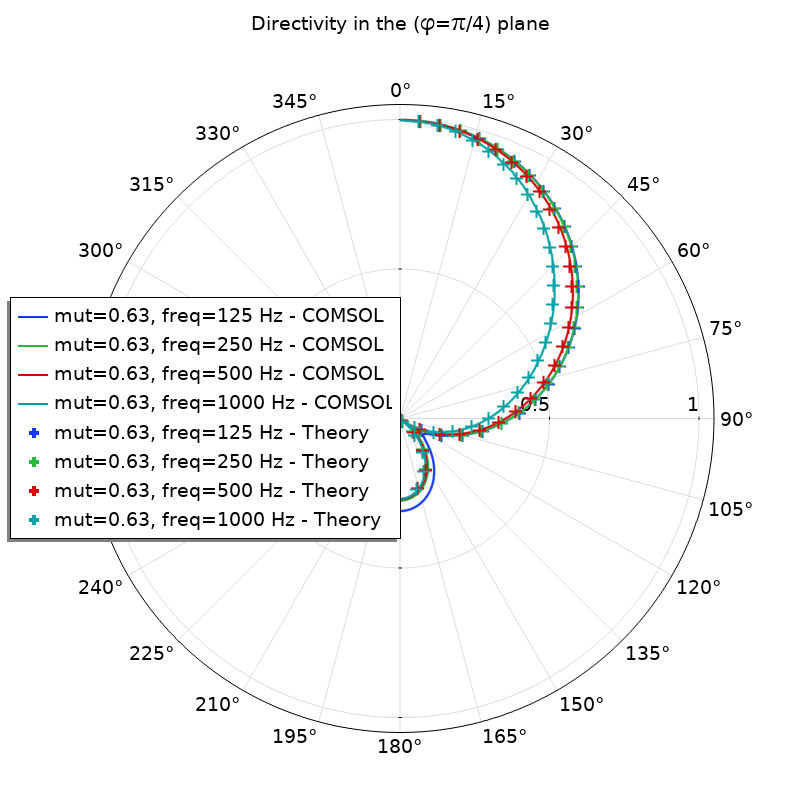}
\caption{}
\end{subfigure}

\begin{subfigure}[h]{0.4\linewidth}
\includegraphics[width=0.9\linewidth]{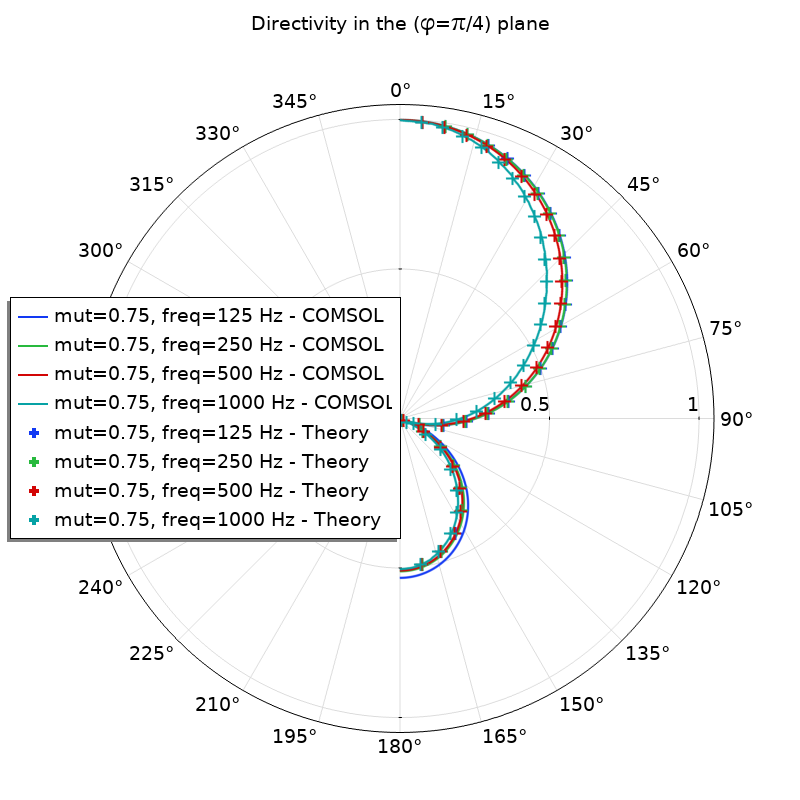}
\caption{}
\end{subfigure}%
\hfill
\begin{subfigure}[h]{0.4\linewidth}
\centering
\includegraphics[width=0.9\linewidth]{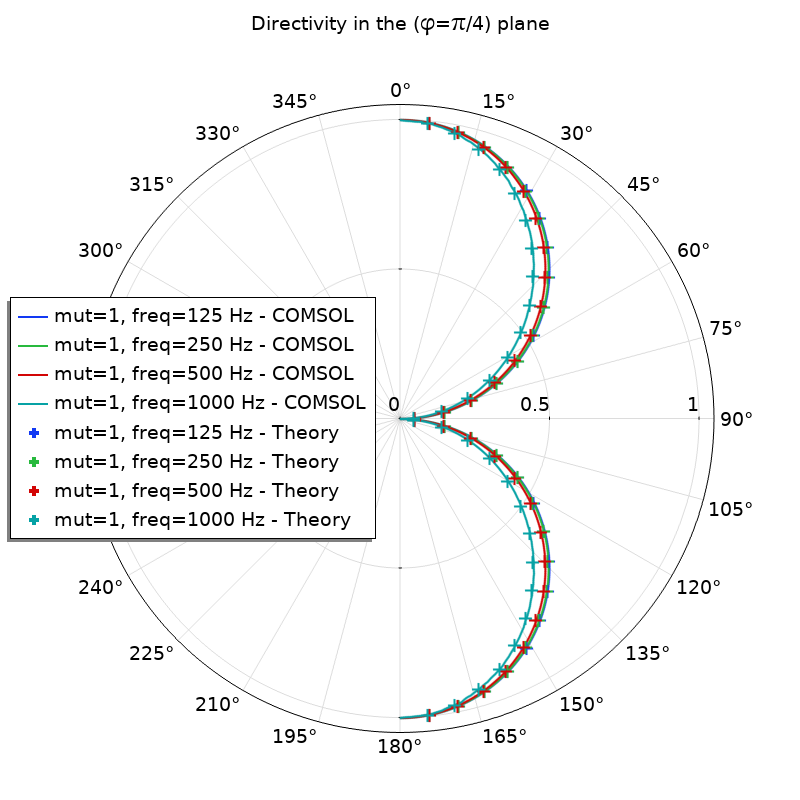}
\caption{}
\end{subfigure}%
\caption{Computed directivity $D(\theta,\frac{\pi}{4})$ in the $\varphi=\frac{\pi}{4}$ plane for each octave band between \SI{125}{Hz} and \SI{1}{kHz} simulated with COMSOL Multiphysics (plain lines) and compared to the theoretical expression of Eq. \eqref{eq:far field dual CDT simplified} (plus markers), for different values of $\mu_t$: (a) omnidirectional ($\mu_t=0$); (b) subcardioid ($\mu_t=0.3$); (c) cardioid ($\mu_t=0.5$); (d) supercardioid ($mu_t=0.63$); (e) hypercardioid ($\mu_t=0.75$); (f) bidirectional ($\mu_t=1$)}
\label{fig:simulation-super and hyper}
\end{figure}

\subsubsection{Pressure along $z$ axis}
\begin{figure}[h!]
\begin{subfigure}[h]{0.4\linewidth}
\centering
\includegraphics[width=0.9\linewidth]{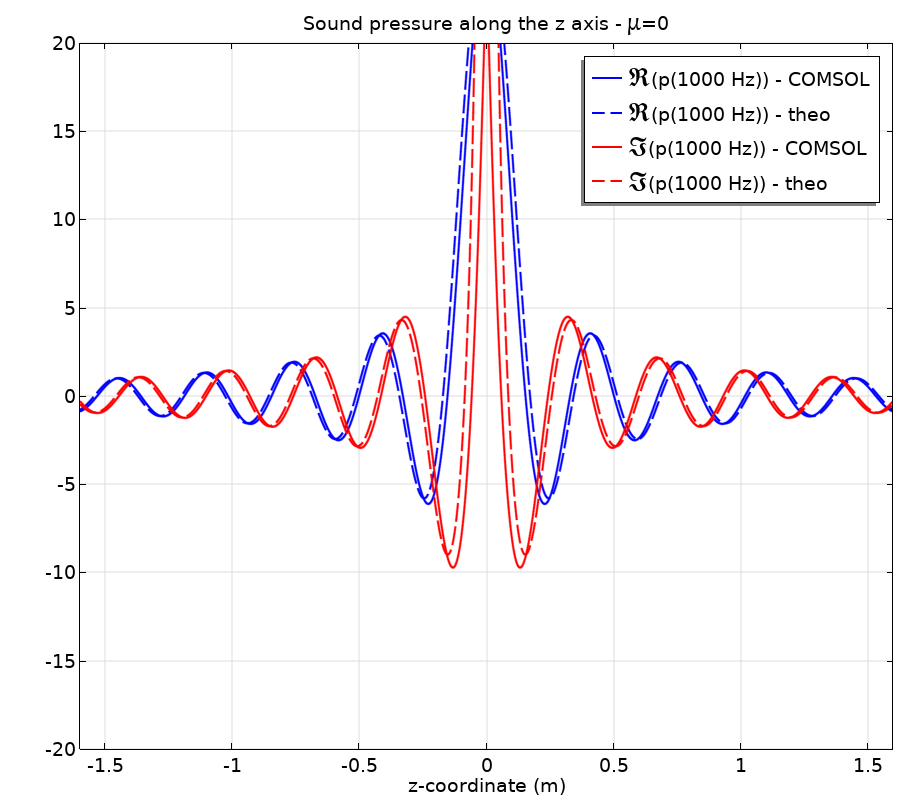}
\caption{}
\end{subfigure}
\hfill
\begin{subfigure}[h]{0.4\linewidth}
\centering
\includegraphics[width=0.9\linewidth]{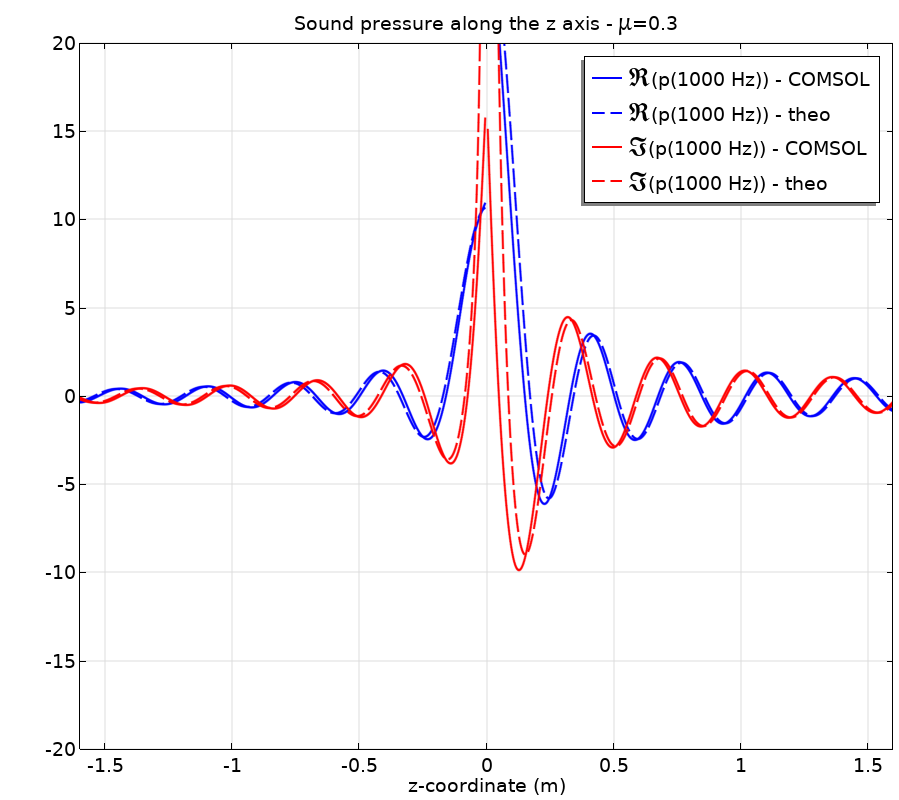}
\caption{}
\end{subfigure}

\begin{subfigure}[h!]{0.4\linewidth}
\centering
\includegraphics[width=0.9\linewidth]{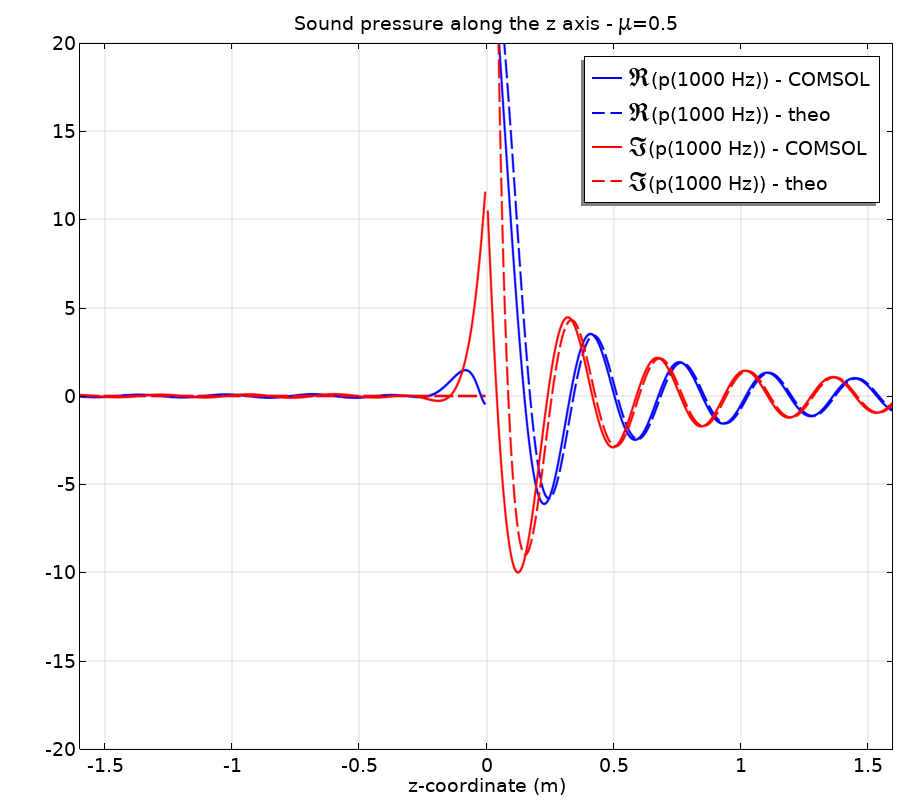}
\caption{}
\end{subfigure}%
\hfill
\begin{subfigure}[h]{0.4\linewidth}
\centering
\includegraphics[width=0.9\linewidth]{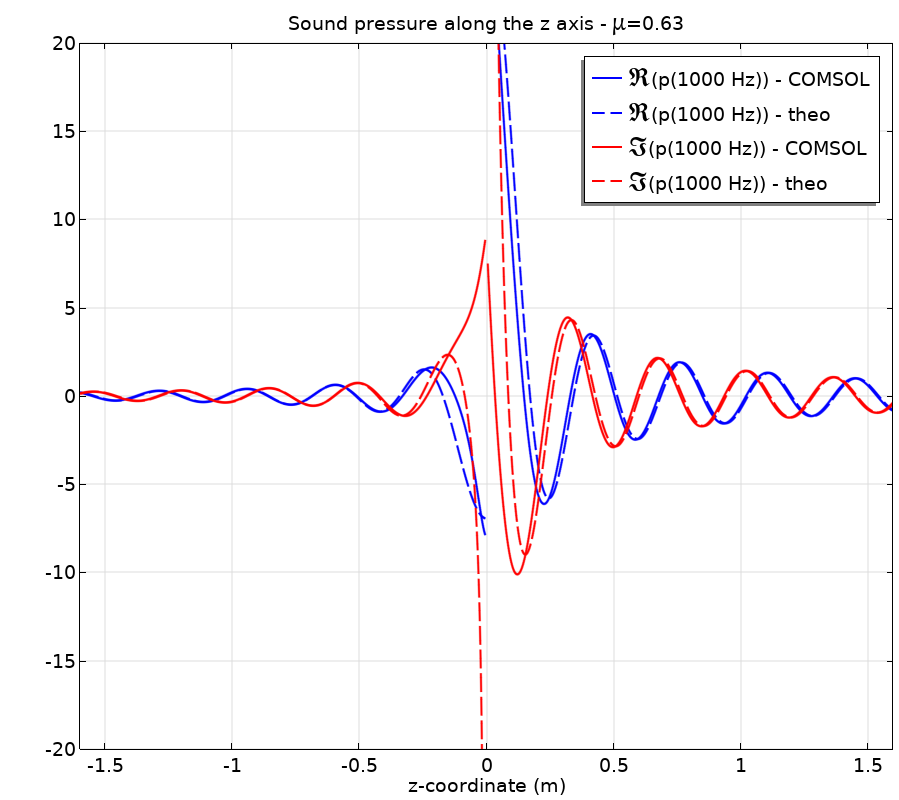}
\caption{}
\end{subfigure}

\begin{subfigure}[h]{0.4\linewidth}
\centering
\includegraphics[width=0.9\linewidth]{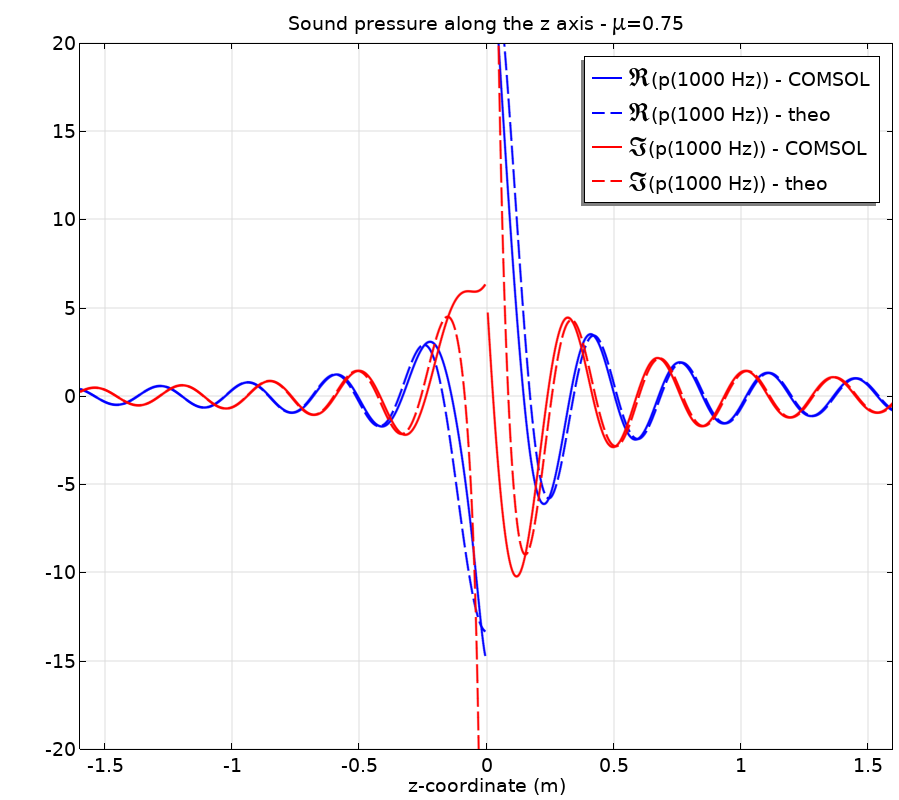}
\caption{}
\end{subfigure}%
\hfill
\begin{subfigure}[h]{0.4\linewidth}
\centering
\includegraphics[width=0.9\linewidth]{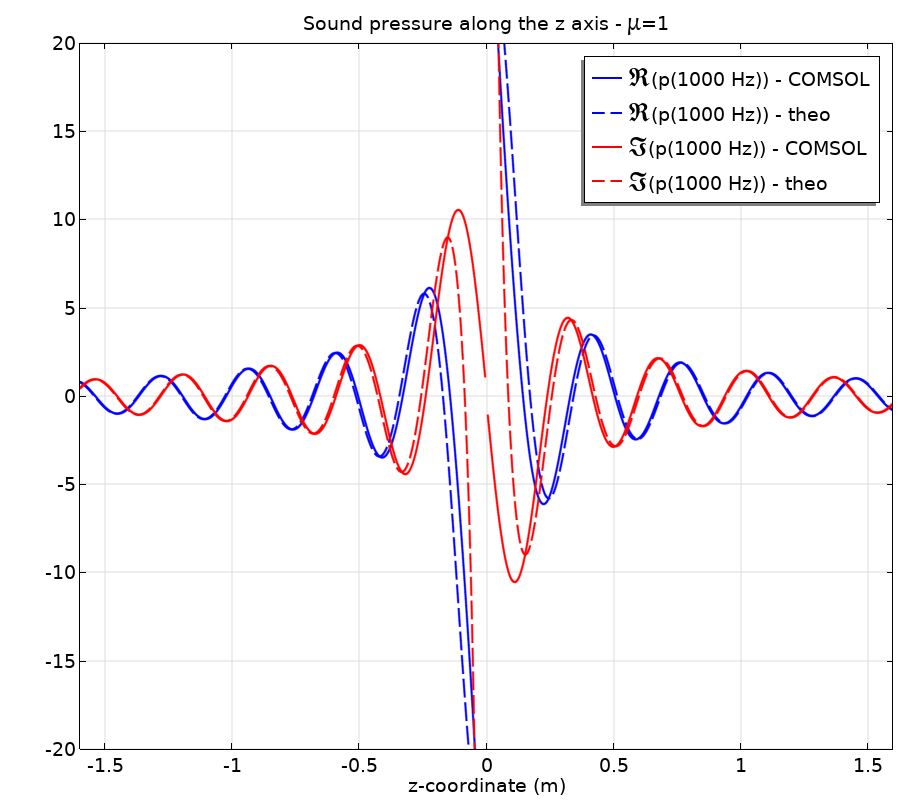}
\caption{}
\end{subfigure}%
\caption{Sound pressure (plain lines: real part; dashed lines: imaginary part) computed along the $z$ axis with COMSOL Multiphysics (blue) and compared with the theoretical formulation of Eq. \eqref{eq:far field dual CDT simplified} at $f=$\SI{1}{kHz} and for different values of $\mu_t$:  (a) omnidirectional ($\mu_t=0$); (b) subcardioid ($\mu_t=0.3$); (c) cardioid ($\mu_t=0.5$); (d) supercardioid ($mu_t=0.63$); (e) hypercardioid ($\mu_t=0.75$); (f) bidirectional ($\mu_t=1$)}
\label{fig:simulation-all directivites- axis}
\end{figure}

To further confirm the validity of the analytical model of the dual CDT arrangement, the pressure as a function of distance along the $z$ axis is also computed. The real part and imaginary part of the total sound pressure $acpr.p_t$ computed with COMSOL over the $z$ axis at each frequency is then compared to the theoretical expression given in Eq. \eqref{eq:far field dual CDT simplified}, also multiplied by the piston directivity of Eq. \eqref{eq:piston directivity}. The results obtained at $f=$ \SI{1}{kHz} are illustrated on Figure \ref{fig:simulation-all directivites- axis} for all cases. We observe a good agreement between the COMSOL full wave simulations and the simple theoretical model elaborated in Sec. \ref{sec:ideal CDTs} in the far field ($kr \gg 1$), validating the analytical model.

The next section will present experimental results obtained on a square dual CDT prototype.

\section{Experimental validation}\label{Sec:exp}

\begin{figure}[h!]
\begin{subfigure}[h]{0.45\linewidth}
\centering
\includegraphics[width=0.9\linewidth]{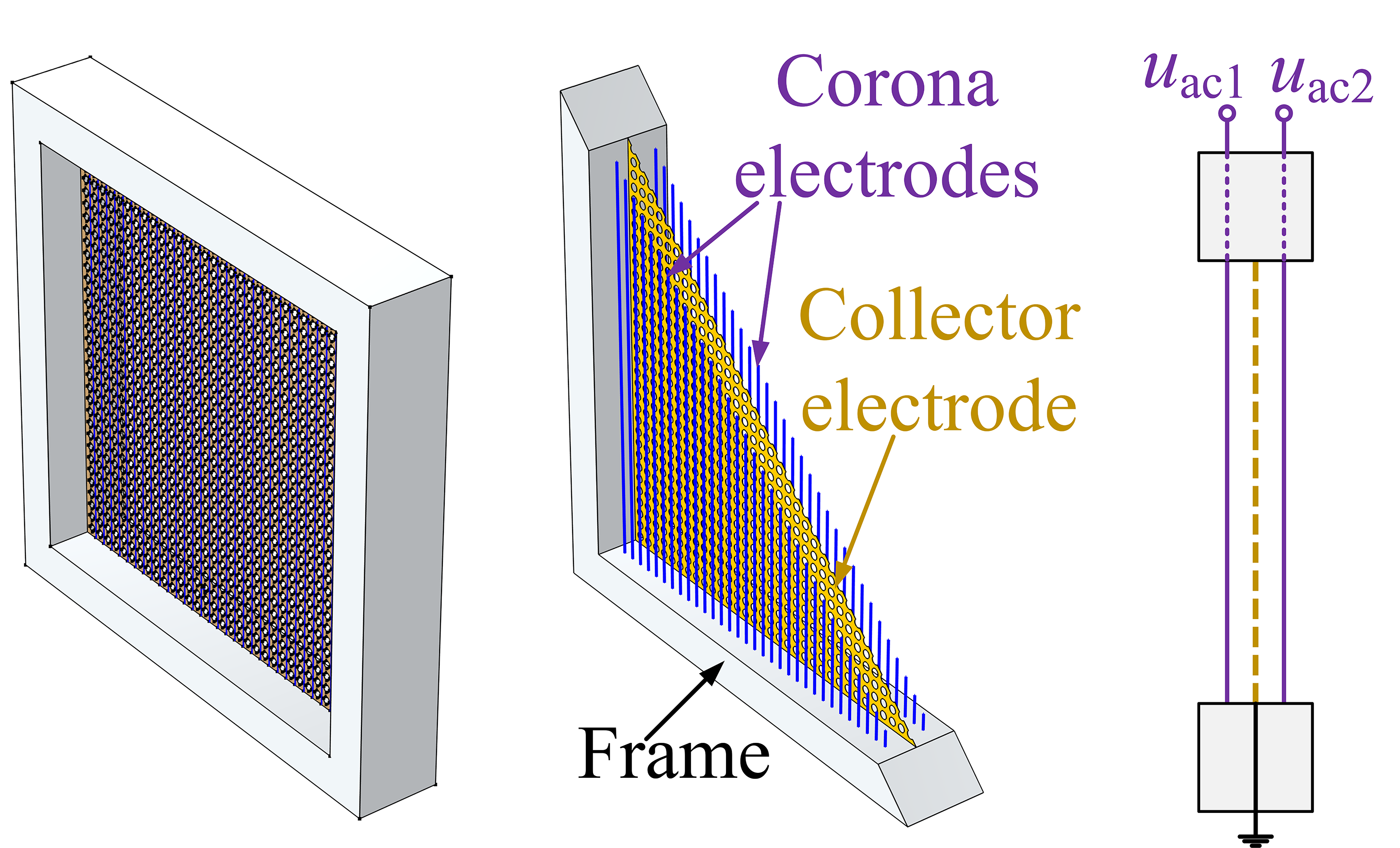}
\caption{Construction principle}
\end{subfigure}
\hfill
\begin{subfigure}[h]{0.25\linewidth}
\centering
\includegraphics[width=\linewidth]{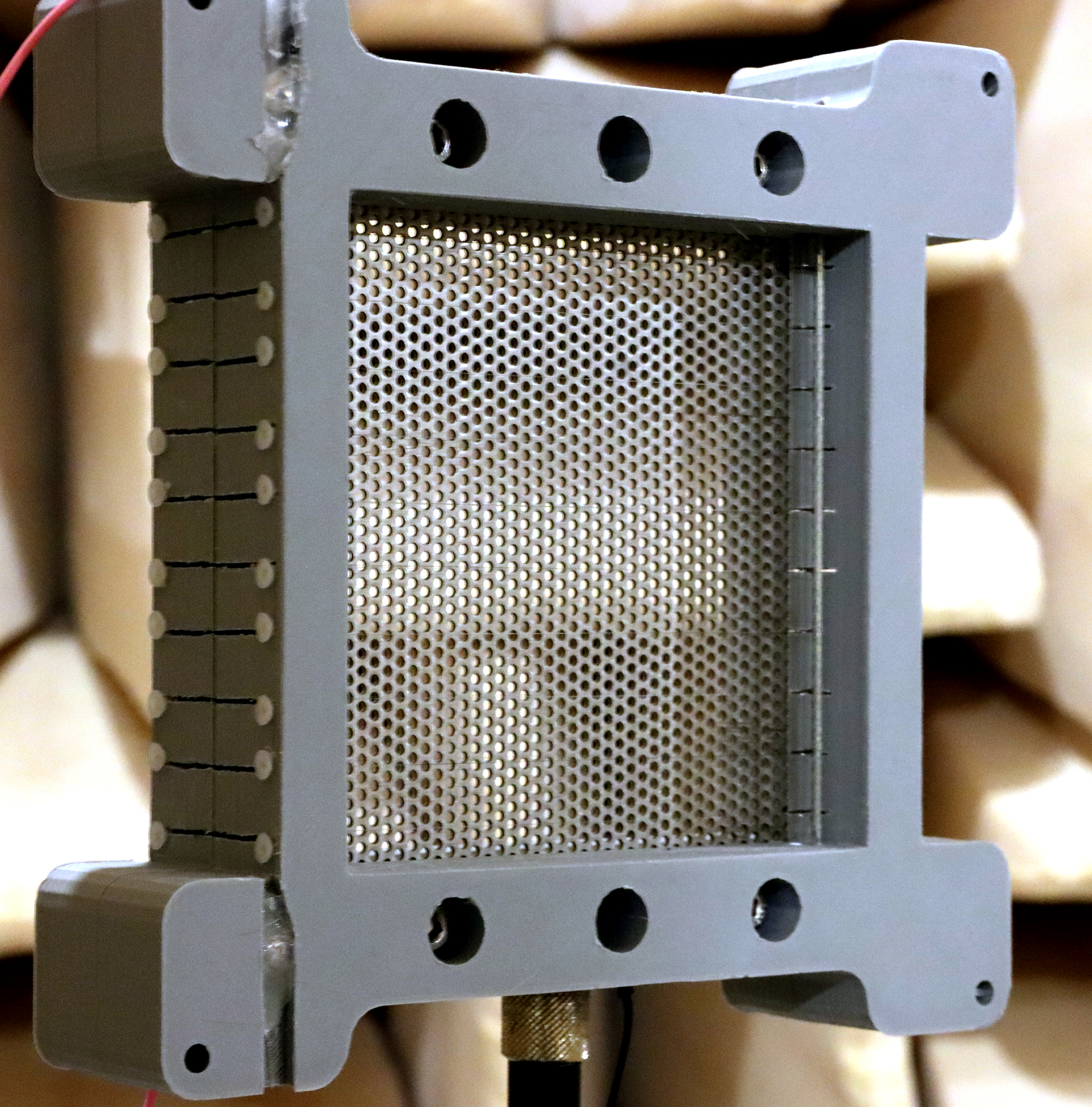}
\caption{Front view}
\end{subfigure}
\hfill
\begin{subfigure}[h]{0.25\linewidth}
\includegraphics[width=\linewidth]{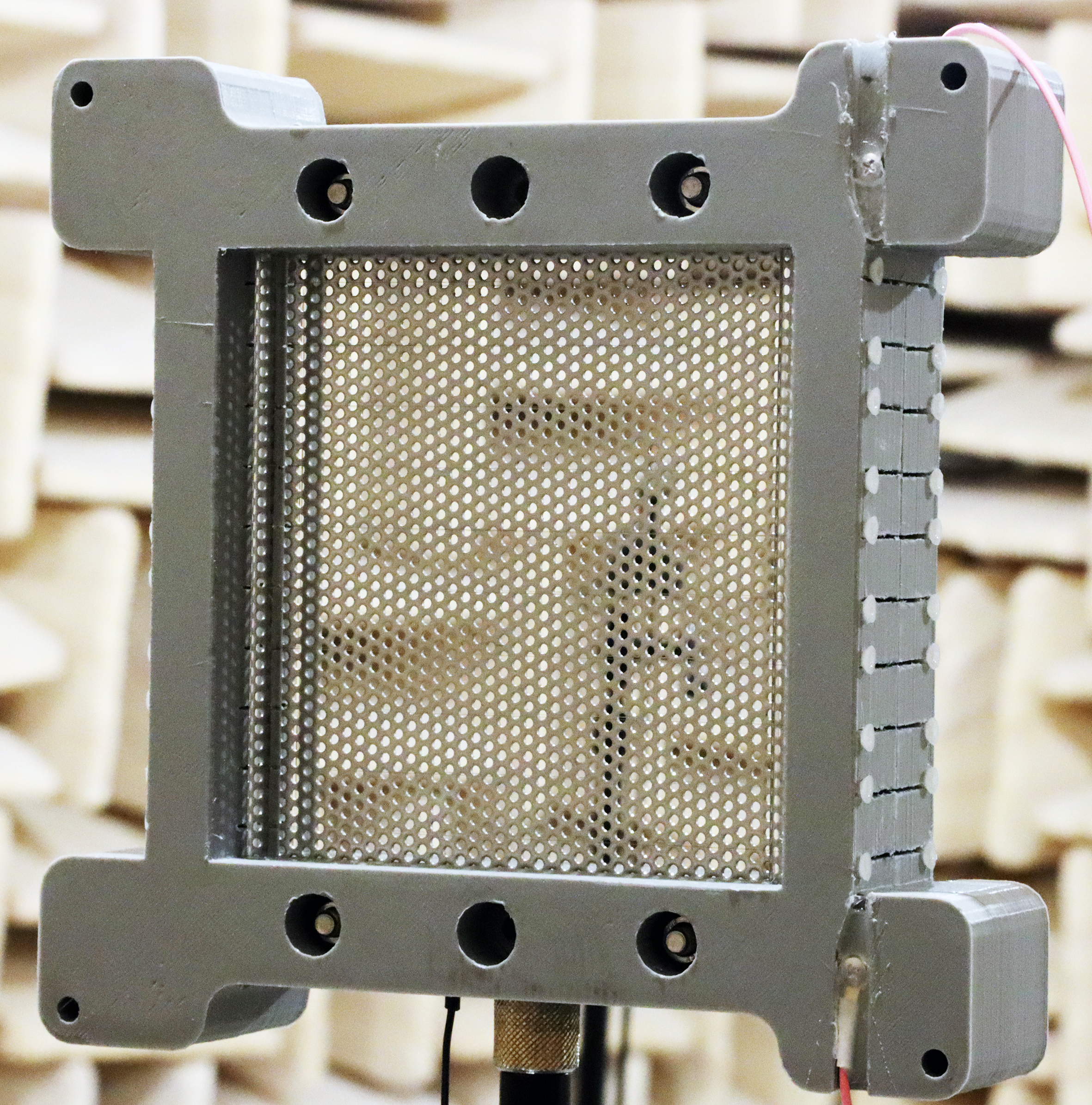}
\caption{Rear view}
\end{subfigure}%
\caption{Dual CDT prototype}
\label{fig:ptototype}
\end{figure}

The square dual CDT is assembled on a common frame, constructed in 3D prototyping. A perforated grid made of Aluminium of thickness $t=$ \SI{1}{mm} is assembled in the median plane of the frame, and two networks of thin Tungsten wires of diameter $d_{wire}=$\SI{100}{\mu m} with inter-wire distance of \SI{11}{mm} are assembled symmetrically on both sides, at a distance $d=$\SI{5}{mm} to the metallic grid. The dual CDT transducer is fed with DC+ac voltages from a custom dual-channel, High Voltage valve-based power amplifier, described in Ref. \cite{Sergeev2022a}. \figref{fig:ptototype} shows the construction principle and a picture of the experimental prototype taken from both sides.

\begin{figure}
    \centering
    \includegraphics[width=0.5\linewidth]{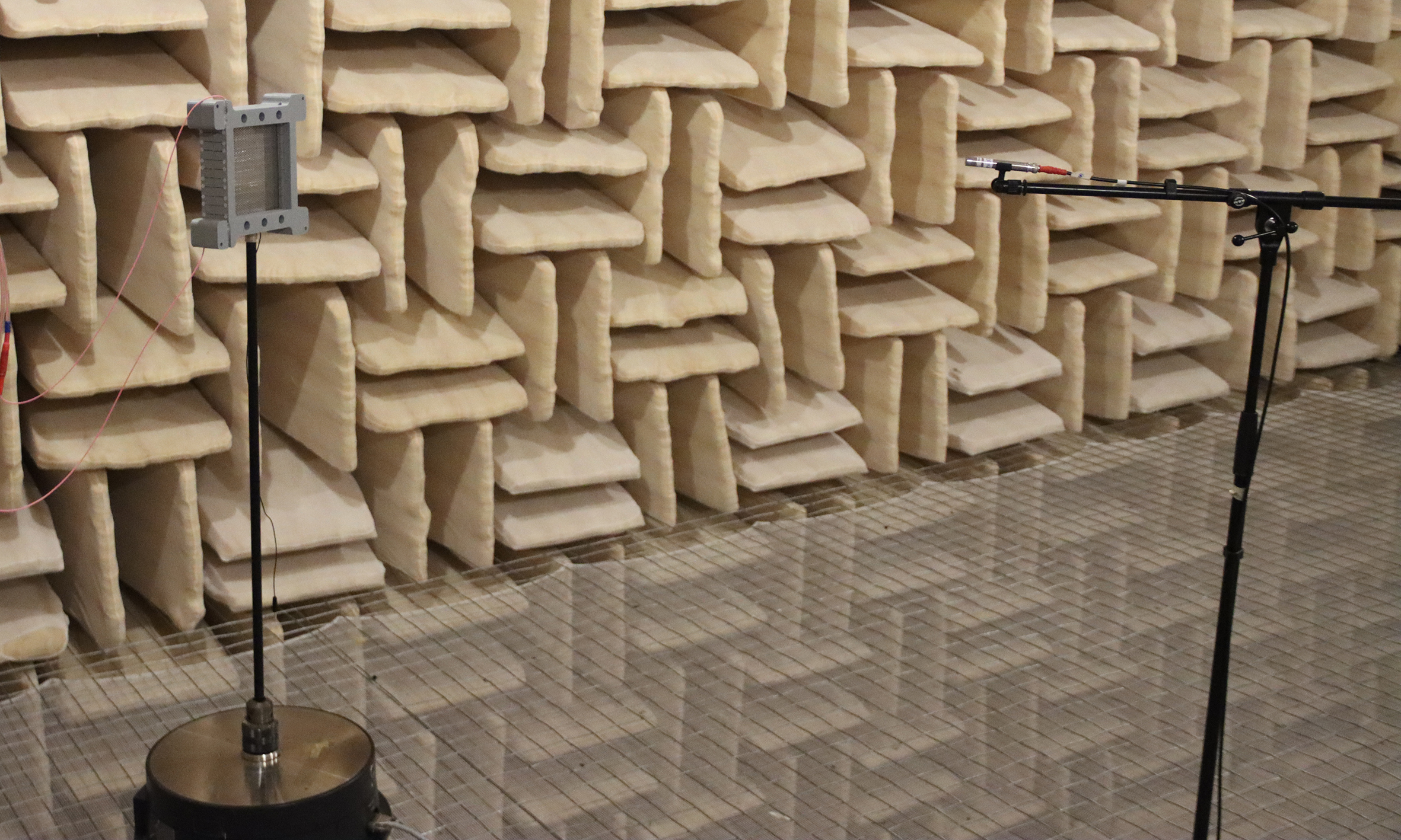}
    \caption{Experimental setup}
    \label{fig:experimental-setup}
\end{figure}

The dual CDT is mounted on a microphone stand fixed to a Bruel \& Kjaer Type 9640 Turntable System, and a PCB 378B02 half-inch microphone is aligned at \SI{1}{m} from the dual CDT center. The signal acquisition and processing is performed with a Bruel \& Kjaer Type 3160 Pulse Multichannel Sound and Vibration Analyzer, each CDT channel on the power amplifier being excited with a synchronous swept sine ranging from \SI{60}{Hz} to \SI{2060}{Hz} with a sweep speed of \SI{10}{mdec/s}, and individual voltages $u_{ac1}$ and $u_{ac2}$. The analysis consists in processing the transfer functions between the sound pressure sensed by the microphone and the output signal $u_{ac1}$ delivered to CDT1 amplification channel (that we set to a fixed value for all directivities). \figref{fig:experimental-setup} illustrates the experimental setup (the power amplifier and cabling are not visible here).

In this experimental evaluation, we will consider unknown CDTs parameters, following the methodology of Sec. \ref{sec:unknown CDTs}.

\subsection{Calibrating the Directional Corona Discharge Loudspeaker: Omnidirectional and bidirectional directivities}\label{sec:exp calibration}
The first step consist in identifying the combination of voltages ($u_{ac1,m},u_{ac2,m}$) that gives rise to a pure omnidirectional directivity, and the pair ($u_{ac1,d},u_{ac2,d}$) that gives rise to a pure bidirectional directivity. To ensure the target directivity is achieved, two measurements of frequency responses are made for each settings:
\begin{itemize}
    \item Measurement at angles $\theta=0$ and $\theta=\frac{\pi}{2}$ for the omnidirectional target, and identification of the value for which both curves are optimally aligned,
    \item Minimization of the frequency response at angle $\theta=\frac{\pi}{2}$ for the bi-directional target.
\end{itemize}

\begin{figure}[ht]
\begin{subfigure}[h]{0.5\linewidth}
\centering
\includegraphics[width=\linewidth]{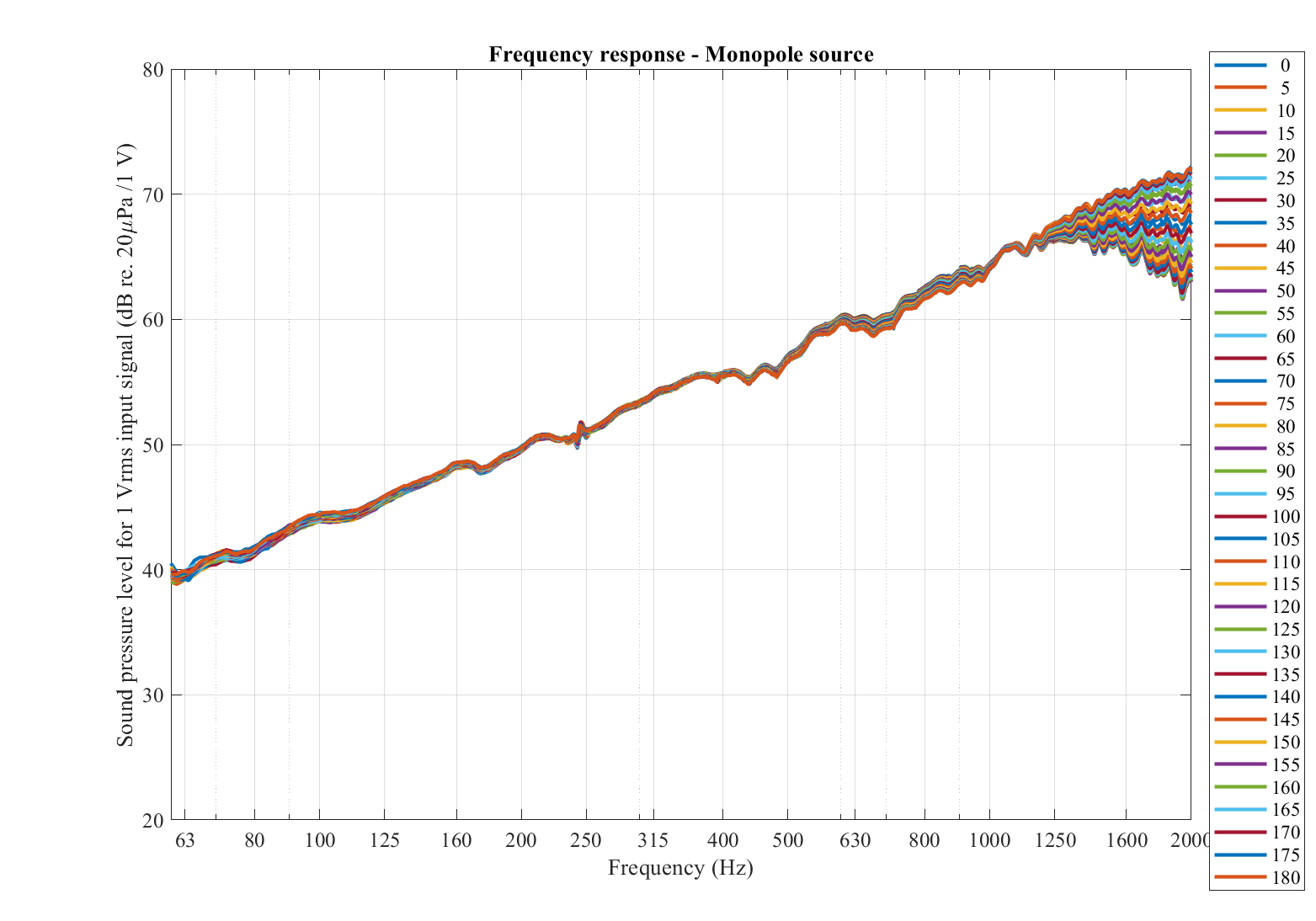}
\caption{}
\end{subfigure}
\hfill
\begin{subfigure}[h]{0.5\linewidth}
\includegraphics[width=\linewidth]{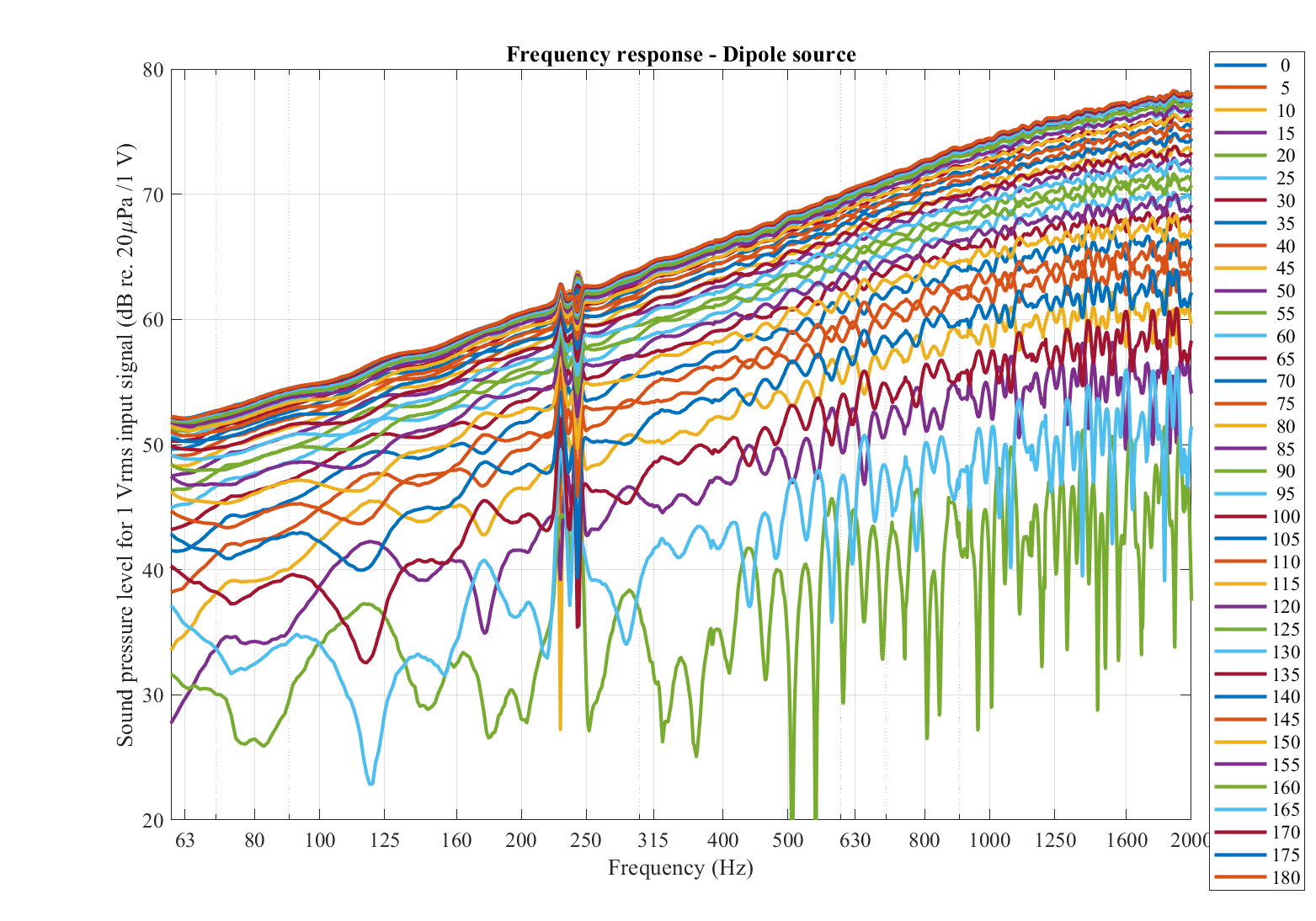}
\caption{}
\end{subfigure}%
\caption{Frequency responses $20\log_{10}(\frac{p(\SI{1}{m},\theta)}{u_{ac1}}.\frac{\SI{1}{V}}{\SI{20}{\mu Pa}})$ measured at different angles for (a) the monopolar and (b) the dipolar settings}
\label{fig:meas-FRF_monopole and dipole}
\end{figure}

\begin{figure}[ht]
\begin{subfigure}[h]{0.5\linewidth}
\centering
\includegraphics[width=\linewidth]{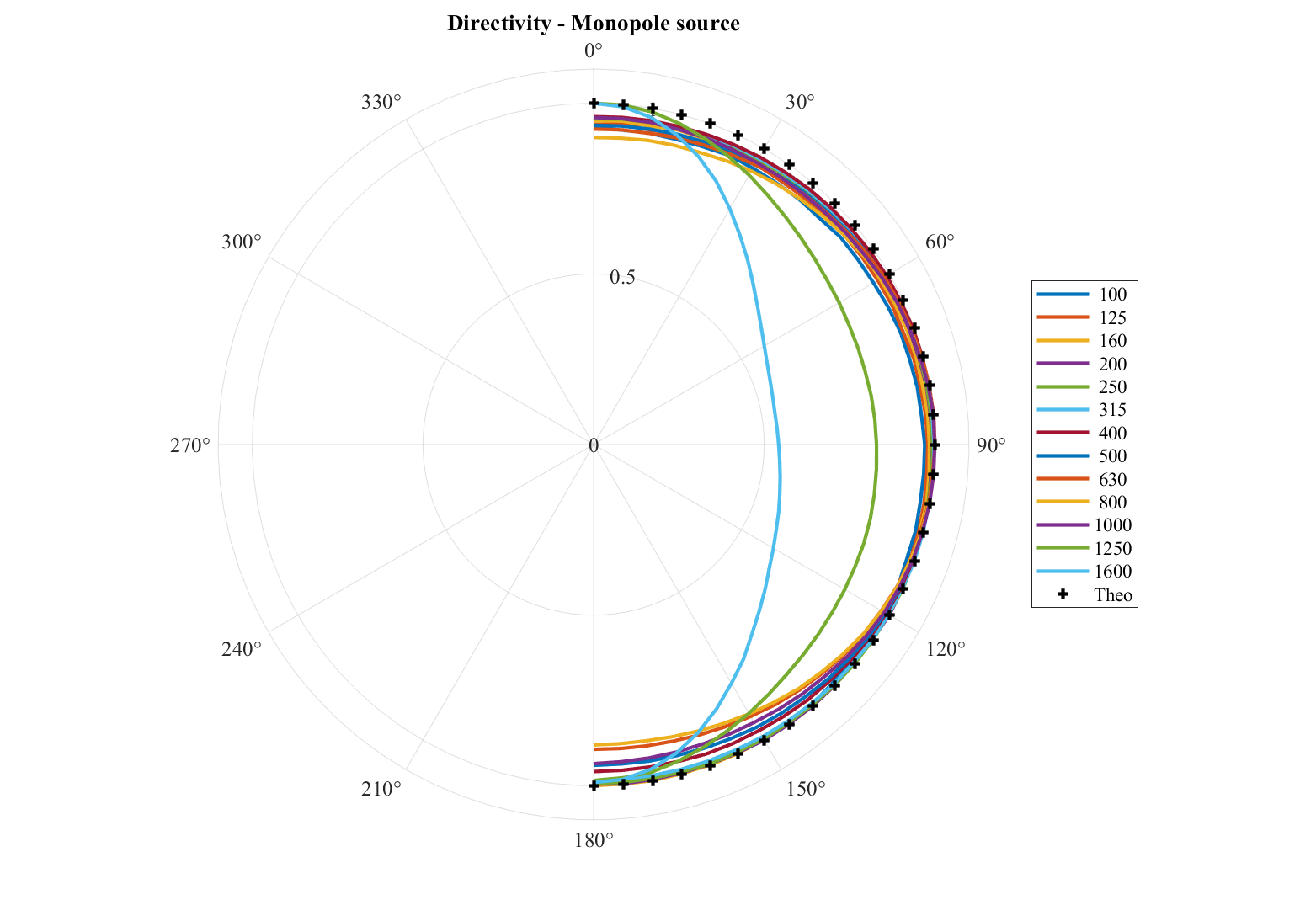}
\caption{}
\end{subfigure}
\hfill
\begin{subfigure}[h]{0.5\linewidth}
\includegraphics[width=\linewidth]{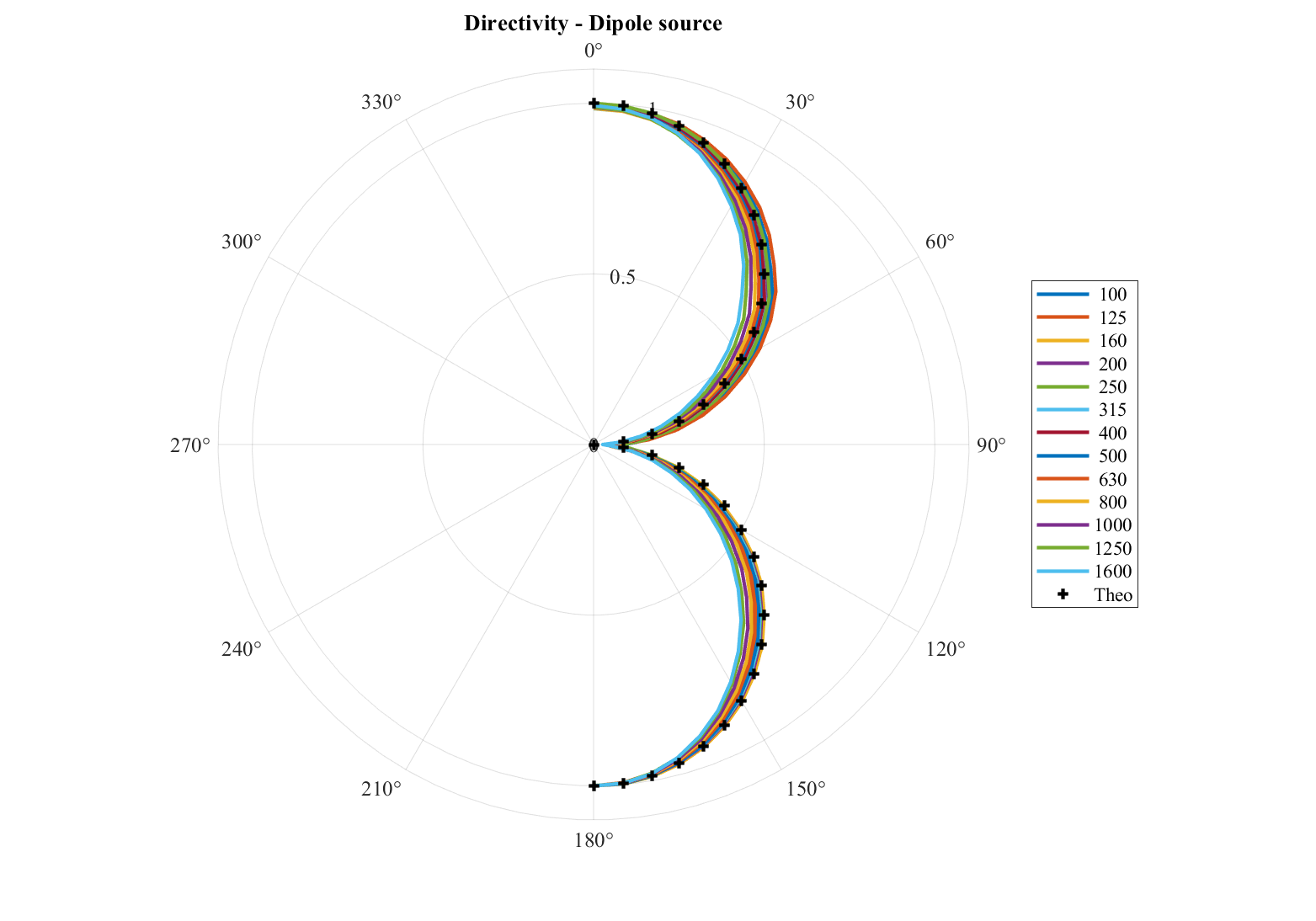}
\caption{}
\end{subfigure}%
\caption{Directivity measured, after the calibration procedure, at different angles for (a) the omnidirectional ($\mu=0$) and (b) the bidirectional ($\mu=1$) settings. For each case, the theoretical directivity $D_t(\theta)=1-\mu_t +\mu_t \cos \theta$ is presented with black plus markers}
\label{fig:meas-directivity_monopole and dipole}
\end{figure}

Once the two voltage pairs are identified, the frequency responses (FRF) are measured for each angle $\theta = n \frac{\pi}{36} $ with integer $n \in [0:36]$. The FRFs for the two set of measurements are given for each angle $\theta$ in \figref{fig:meas-FRF_monopole and dipole}, and the corresponding directivity sampled at each third-octave central frequencies are illustrated in \figref{fig:meas-directivity_monopole and dipole}. The results reflect the calibration procedure, which allows determining the reference voltages pairs ($u_{ac1,m},u_{ac2,m}$) and ($u_{ac1,d},u_{ac2,d}$). The observed agreement with the theoretical results obtained from COMSOL simulations demonstrates the precision of the procedure up to \SI{1}{kHz}.

\subsection{Achieving unidirectional directivities}

\begin{figure}[ht]
\begin{subfigure}[h]{0.5\linewidth}
\centering
\includegraphics[width=\linewidth]{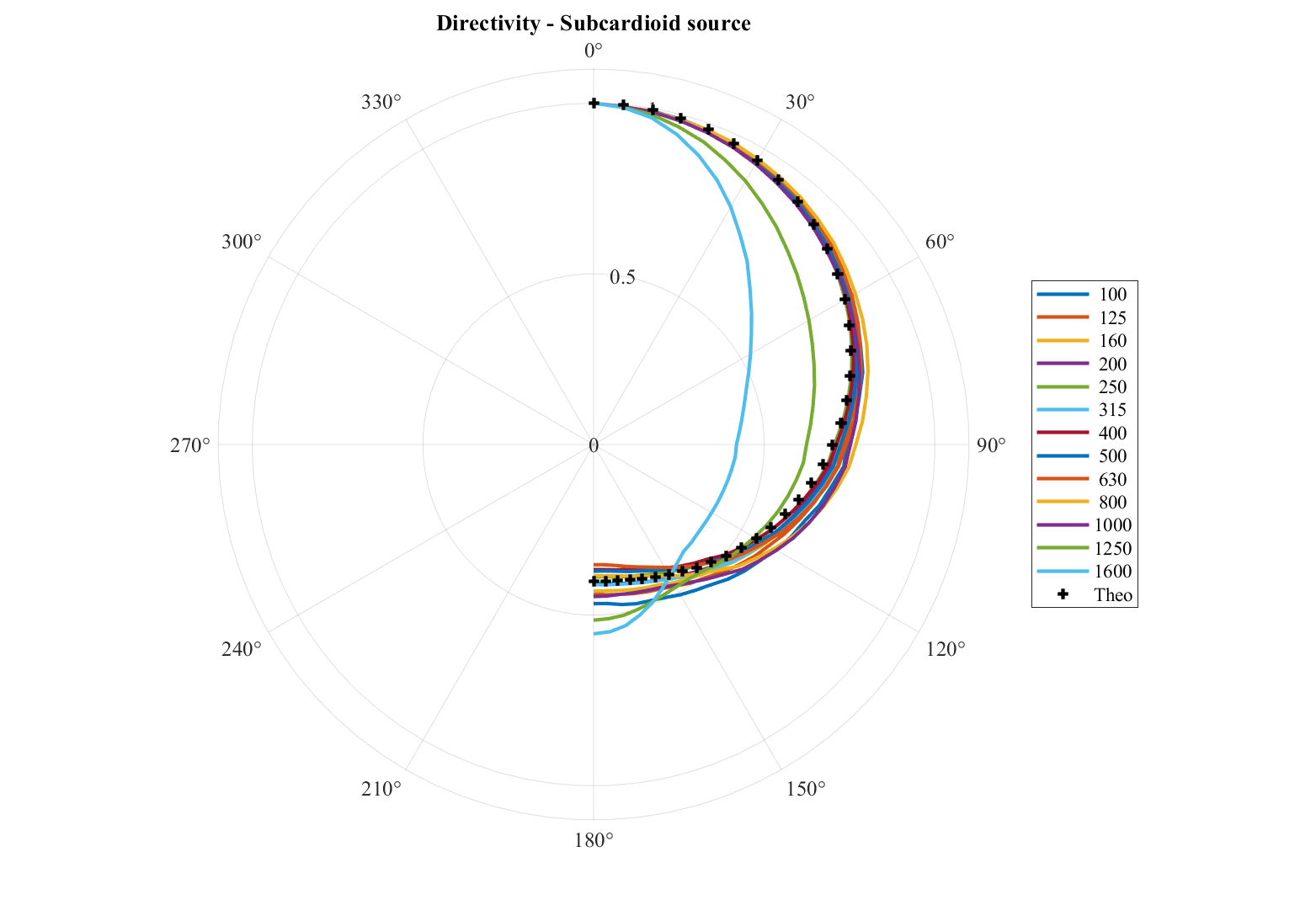}
\caption{}
\end{subfigure}
\hfill
\begin{subfigure}[h]{0.5\linewidth}
\includegraphics[width=\linewidth]{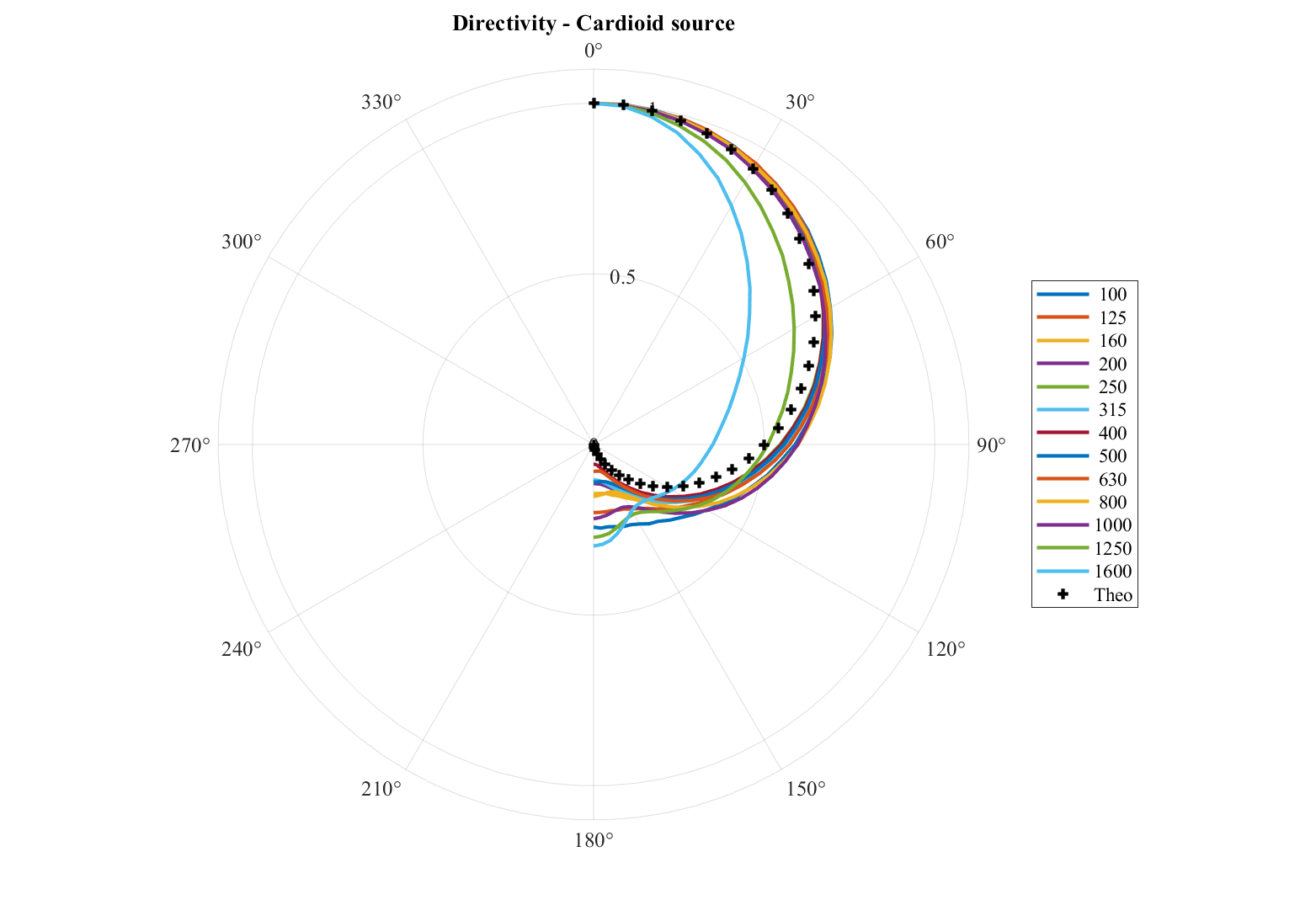}
\caption{}
\end{subfigure}%

\begin{subfigure}[h]{0.5\linewidth}
\centering
\includegraphics[width=\linewidth]{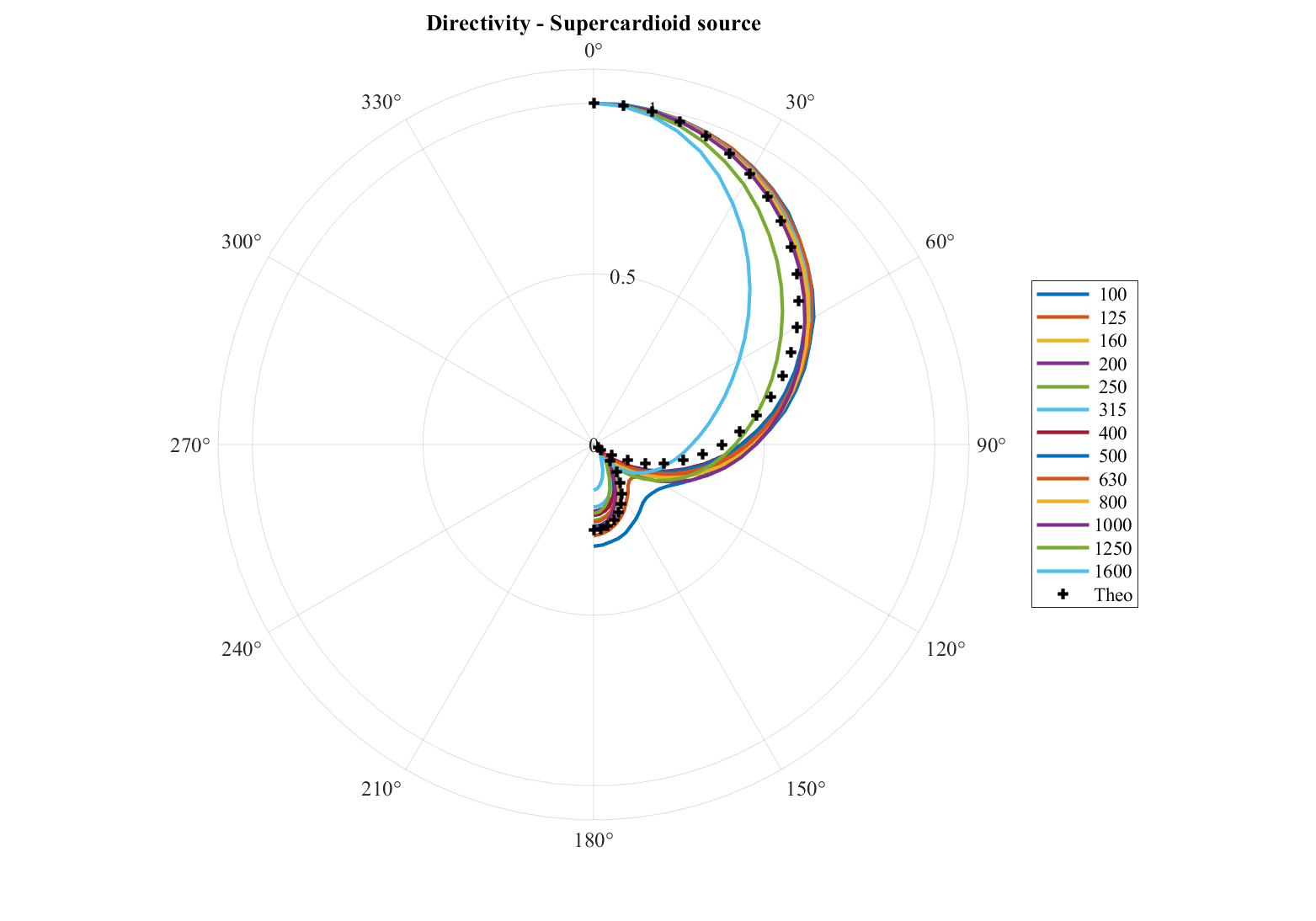}
\caption{}
\end{subfigure}
\hfill
\begin{subfigure}[h]{0.5\linewidth}
\includegraphics[width=\linewidth]{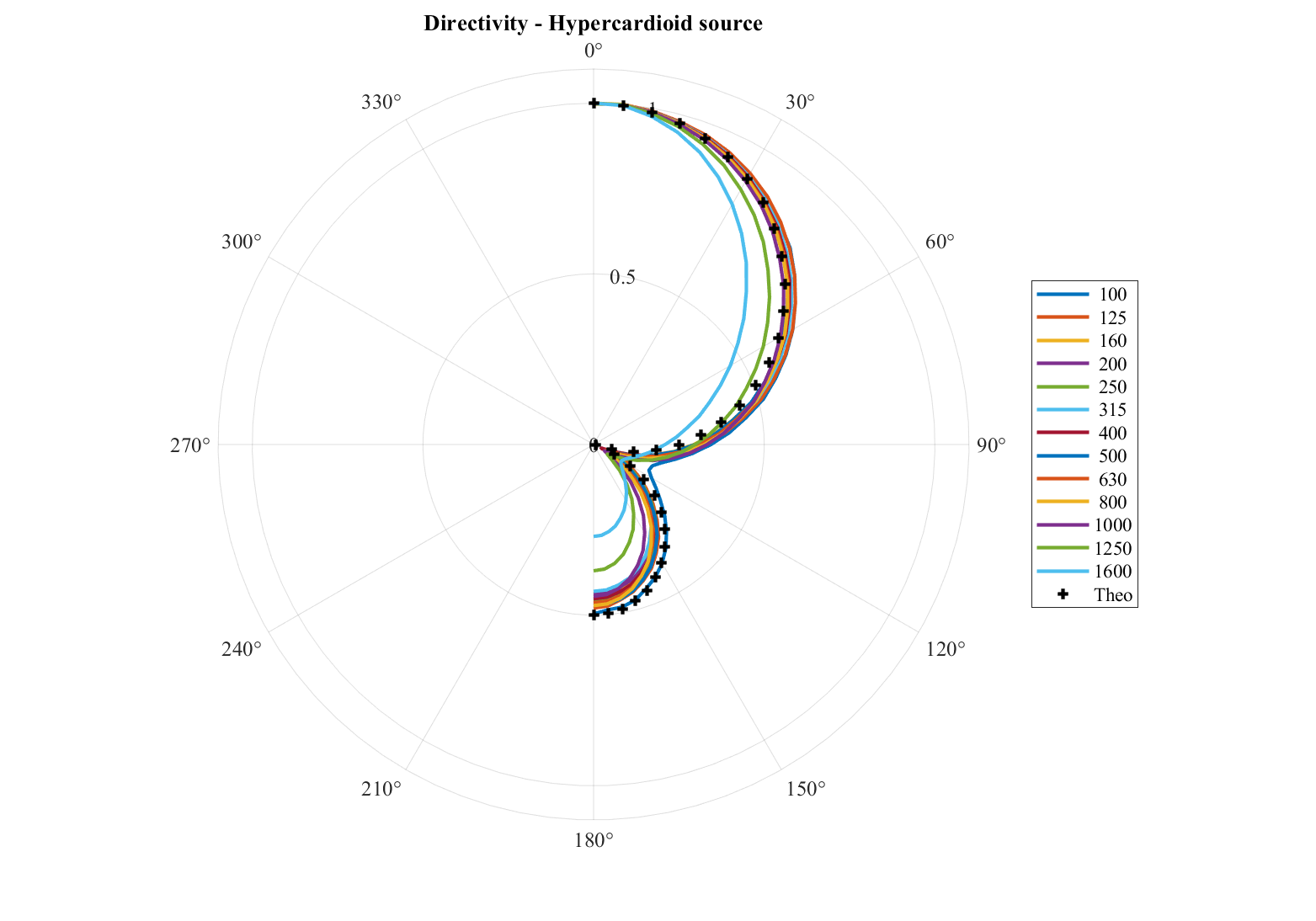}
\caption{}
\end{subfigure}%
\caption{Directivity measured at different angles for:  (a): the subcardioid ($\mu=0.63$); (b) the cardioid ($\mu=0.5$); (c); the supercardioid ($\mu=0.63$); and (d) the hypercardioid ($\mu=0.75$) settings. For each case, the theoretical directivity $D(\theta)=1-\mu +\mu \cos \theta$ is presented with black plus markers}
\label{fig:meas-directivity_unidirectional}
\end{figure}

Once the voltages pairs corresponding to $\mu_t=0$ and $\mu_t=1$ have been identified, the other pairs have been derived corresponding to any of the targeted $\mu_t$, following the methodology detailed in Sec. \ref{sec:unknown CDTs}. The directivities achieved with the corresponding pairs of voltages $(u_{ac1,\mu_t},u_{ac2,\mu_t})$ for $\mu_t \in \{0;0.3;0.5;0.63;0.75;1\}$  are illustrated on Figure \ref{fig:meas-directivity_unidirectional}. We can see the perfect match between the achieved directivities and the theoretical ones corresponding to each value of $\mu_t$, at least up to \SI{1}{kHz} above which the directivity of the square piston begins to enter into play and modifies the overall directivity.

\section{Conclusion}
The control over directivity on (conventional) loudspeakers is known to be limited to low-frequencies due to their bulkiness. 
Given the compactness of the CDTs, it is possible to achieve very precise directivities in theory up to the range of a few kHz following the proposed dual CDT concept. In practice, however, the dimensions of the radiator induces a frequency limit above which the impact of planar geometry enters into play. Our results show that the introduced mechanism retains its efficiency below that frequency threshold, significantly outperforming loudspeakers of equal surface in realizing the targeted directivities.

The dual CDT concept has direct applications to directivity control on loudspeakers, owing to its direct conversion of electric voltage to acoustic flow velocity through the (geometry-dependent) radiation impedance. Besides, the dual CDT will be an important asset as secondary source in ducted (1D) active noise reduction systems (eg. active exhaust noise reduction). Indeed the capability to separate the net sound sources on each sides allows for controlling independently the up- and down-stream of the active control source, and thus have a full control over both sound fields individually. By extension, the dual CDT will have direct application to active acoustic metamaterials and metasurfaces, as the control of monopolar/dipolar elements is at the heart of non-reciprocal concepts such as the Willis coupling effect.


\bibliographystyle{ieeetr}
\bibliography{biblio}

\end{document}